\documentclass[usenatbib]{mnras}
\usepackage[T1]{fontenc}
\usepackage{ae,aecompl}

\usepackage{graphicx}	% Including figure files
\usepackage{amsmath}	% Advanced maths commands
\usepackage{amssymb}	% Extra maths symbols
\usepackage{tikz}

\newcommand*\mycirc[1]{%
  \begin{tikzpicture}[baseline=(C.base)]
    \node[draw,circle,inner sep=1pt](C) {#1};
  \end{tikzpicture}}

\title[Photoionisation of Be-like and Li-like atomic  oxygen]
        {{\it K}-shell photoionisation of  O$^{4+}$ and O$^{5+}$   
                       ions : experiment and theory}
\author[B.-M. McLaughlin et al]{
        B. M. McLaughlin$^{1,2}$
        \thanks{bmclaughlin899@btinternet.com :corresponding author},
        J.-M. Bizau$^{3,4}$\thanks{jean-marc.bizau@u-psud.fr},
	    D. Cubaynes$^{3,4}$\thanks{denis.cubaynes@u-psud.fr},
	    S. Guilbaud$^{3}$\thanks{segolene.guilbaud@u-psud.fr},
	    S. Douix$^{4}$\thanks{suzie.douix@synchrotron-soleil.fr},
\newauthor{M. M. Al Shorman$^{5}$\thanks{mmaa1984@yahoo.com}, 
           M. O. A. El Ghazaly$^{6}$\thanks{mael46@gmail.com}, 
           I. Sakho$^{7}$\thanks{animafatima{\_}sakho@yahoo.fr} and
           M. F. Gharaibeh$^{8}$
           \thanks{mgharaibeh@qu.edu.qa: prior address,
                  Department of Physics, 
                  Jordan University of Science and Technology, 
                  Irbid 22110, Jordan}}
\\
$^{1}$Centre for Theoretical Atomic and Molecular Physics (CTAMOP),
          School of Mathematics and Physics,\\
	 Queen's University Belfast, Belfast BT7 1NN, Northern Ireland, UK\\
$^{2}$Institute for Theoretical Atomic and Molecular Physics (ITAMP)\\
	Harvard Smithsonian Center for Astrophysics, 
	 MS-14, Cambridge, MA 02138, USA\\
$^{3}$Institut des Sciences Mol\'{e}culaires d'Orsay (ISMO), 
             CNRS UMR 8214,\\ 
             Univ. Paris-Sud, Universit\'{e} Paris-Saclay, 
             F-91405 Orsay cedex, France\\
$^{4}$Synchrotron SOLEIL - L'Orme des Merisiers, 
            Saint-Aubin - BP 48 91192 Gif-sur-Yvette cedex, France\\
$^{5}$Applied Physics Department, Faculty of Science, Tafila Technical University,
             Tafila 66110, Jordan\\
$^{6}$Astrophysics and Space Sciences Section, 
            Jet Propulsion Laboratory, Caltech, Pasadena, CA 91109, USA\\             
$^{7}$Department of Physics, UFR of Sciences and Technologies, 
            University Assane Seck of Ziguinchor, Ziguinchor, Senegal\\
$^{8}$Department of Mathematics, Statistics and Physics, 
 	    P.O. Box 2713, Qatar University, Doha, Qatar\\
}

\date{Accepted --; Received \today; in original form September, 7, 2016}
\pubyear{2016}

\begin{document}

\label{firstpage}
\pagerange{\pageref{firstpage}--\pageref{lastpage}}
\maketitle

\begin{abstract}
 {Absolute cross sections for the {\it K}-shell photoionisation 
 of  Be-like  (O$^{4+}$) and Li-like (O$^{5+}$) atomic oxygen ions 
 were measured for the first time (in their respective {\it K}-shell regions) by
 employing the ion-photon merged-beam technique at the SOLEIL 
 synchrotron-radiation facility in Saint-Aubin, France.  
 High-resolution spectroscopy with 
 E/$\Delta$E $\approx$ 3200 ($\approx$ 170 meV, FWHM) 
 was achieved with photon energy from 550 eV up to 670 eV.
 Rich resonance structure observed  
 in the experimental spectra is analysed using the 
 R-matrix with pseudo-states (RMPS) method.  Results are also 
 compared with the screening constant by unit nuclear charge (SCUNC) calculations. 
 We  characterise and identify the strong
  $\rm 1s \rightarrow 2p$ resonances  for both ions and the 
 weaker $\rm 1s \rightarrow  np$  resonances ($ n \ge 3$)
 observed in the {\it K}-shell spectra of O$^{4+}$.}
\end{abstract}


%%%%%%%%%%%%%%%%% BODY OF PAPER %%%%%%%%%%%%%%%%%%

\section{Introduction}
The launch of the satellite Astro-H (re-named Hitomi) on February 17, 2016,
was expected to provide x-ray spectra of unprecedented quality and  
would have required a wealth of atomic and molecular data 
on a range of collision processes to assist with 
 the analysis of spectra from a variety of astrophysical objects.
 The subsequent break-up on March 28, 2016 of Hitomi now leaves 
 a void in observational x-ray spectroscopy. 
In the intervening period before the launch of the next x-ray satellite 
mission, ground-based x-ray experiments along with 
theoretical cross sections will continue to 
be benchmarked against one and other, in order to extend our database of 
knowledge on astrophysical important ions of carbon, nitrogen and oxygen.  
 
Measurements of cross sections for photoionisation of atoms and ions are
essential data for testing theoretical methods in fundamental
atomic physics \citep{berko1979,West2001,Kjeldsen2002,Mueller2015b} 
and for modeling of many physical
systems, for example terrestrial plasmas, the upper atmosphere, 
and a broad range of astrophysical objects (quasar
stellar objects, the atmosphere of hot stars, proto-planetary
nebulae, H II regions, novae, and supernovae)
\citep{Lee2001,Blustin2002,Blustin2003,Kaastra2002,Kaastra2004, Juett2004,Pinto2013,Pinto2014,Nicastro2016a,Nicastro2016b}. 

X-ray spectroscopy of highly ionised atomic oxygen is used to probe the hot 
 gaseous halo of the Milky Way \citep{Nicastro2016c,Miller2015,Gupta2012,Gupta2014}.
Multiply ionisation stages of C, N, O, Ne and Fe have been observed 
in the ionised outflow in the planetary nebulae NGC 4051, and
measured with the satellite {\it XMM-Newton} \citep{Ogle2004,Pinto2013} 
in the soft-x-ray  region. Low ionised stages of C, N and O have also been used 
in the modelling of x-ray emission from OB super-giants \citep{Cassinelli1981}.
Multiply ionisation stages of O and Fe are also seen in the {\it XMM-Newton} spectra
 from the Seyfert galaxy NGC 3783, including UV imaging, x-ray and UV light curves, 
 the 0.2 -- 10 keV x-ray  continuum, the iron {\it K} - emission line, 
and high-resolution spectroscopy in the modelling 
of the soft x-ray warm absorber
 \citep{Blustin2002, Blustin2003, Krongold2003,Mendoza2012,Gorczyca2013, Gatuzz2013a,Gatuzz2014}.
 
Limited wavelength observations for x-ray transitions 
were recently made on atomic oxygen,
neon and magnesium and their ions with 
the High Energy Transmission Grating (HETG) 
on board the {\it Chandra}  satellite \citep{Liao2013}.  
Strong absorption {\it K}-shell lines
of atomic oxygen, in its various forms of ionisation 
have been observed by the XMM-Newton satellite 
in the interstellar medium, through x-ray 
spectroscopy of low-mass x-ray binaries \citep{Pinto2013}. 
The Chandra and XMM-Newton satellite observations 
may be used to identify absorption 
features in astrophysical sources, 
such as AGN and x-ray binaries and for assistance in  
benchmarking theoretical and experimental work 
\citep{Stolte2013,Oxygen2013, Gorczyca2013, Gatuzz2013a,Gatuzz2013b,Gatuzz2014,Soleil2014,Bizau2015}.

To our knowledge previous experimental 
and observational x-ray lines on Be-like and 
Li-like atomic oxygen ions are 
limited to the $\rm 1s \rightarrow np$ 
 regions (where n=2, 3, 4, and 5), i.e., 
 the $K_{\alpha}$, $K_{\beta}$, $K_{\gamma}$ and $K_{\delta}$ lines,
 in the vicinity of the {\it K}-shell region 
\citep{Tondello1977,Bruch1979,Bruch1987,Hoffmann1990,
       Hoffmann1991,Behar2003,Schmidt2004,Gu2005,Lee2001,Kaastra2002,
       Kaastra2004,Yao2009,Mendoza2012,Ramirez2013,Liao2013,
       Pinto2013,Pinto2014,cabot2013,
       Nicastro2016a,Nicastro2016b}.   
Prior to the present {\it K}-shell investigations 
at the SOLEIL radiation facility on 
photoionisation cross sections and Auger resonance 
states along the atomic oxygen iso-nuclear sequence, 
few experiments have been devoted to the study of 
{\it K}-shell photoionisation of oxygen ions. 
Nicolosi and Tondello \citep{Tondello1977} 
observed satellite lines of He-like and Li-like laser produced 
plasmas of Be, B, C, N and O.  Auger spectra of  core-excited oxygen ions 
emitted in the collision of fast oxygen-ion beams with gas targets and foils 
were measured by Bruch and co-workers \citep{Bruch1979,Bruch1987}. 
Hoffmann and co-workers \citep{Hoffmann1990,Hoffmann1991}    
measured Auger resonance energies in electron-impact 
ionisation studies of Be-like and Li-like ions.
{\it K}-shell x-ray lines from inner-shell excited and ionised ions of oxygen, 
were observed using the Lawrence  Livermore National 
Laboratory EBIT from O$^{2+}$ to O$^{5+}$ \citep{Schmidt2004,Gu2005}. 

For the oxygen iso-nuclear sequence theoretical data are available for the 
energies for {\it K}-shell Auger or radiative transitions.
Resonance energies and line widths for Auger transitions 
in Be-like \citep{Bruch1979} and Li-like atomic ions \citep{Nicolaides1993} 
have been calculated  using a variety of methods  
such as 1/Z perturbation theory \citep{Safronova2002}, 
multi-configuration Dirac Fock (MCDF) 
\citep{Bruch1979,Chen1985,Chen1986,Chen2006}, 
the  saddle-point-method (SPM) with  R-matrix \citep{Davis1985,Davis1989}  
and complex-coordinate rotation methods \citep{Yeager2012a,Yeager2012b}. 
We note that Moore \citep{Moore1993}  made an assessment of the energy levels 
of Carbon, Nitrogen and Oxygen atoms and their ions. 
Chen and co-workers calculated Auger and radiative decay of $1s$-vacancy states in the  
Be iso-electronic sequence using the MCDF approach \citep{Chen1985,Chen1986,CC1987}. 
Zhang and Yeager \citep{Yeager2012a} used 
the SPM with rotation  \citep{Davis1985, BW2000}
to calculate energy levels and Auger decay widths for the 
$\rm 1s2s^22p\;^{1,3}P^o$   levels in Be-like system.  
The recent saddle-point with rotation calculations by Yeager and co-workers 
\citep{Yeager2012a}, for  resonance energies and Auger decay rates, 
for the $\rm 1s2s^22p\;^{1}P^o$ 
 levels in Be-like carbon showed excellent agreement 
with previous measurements made at the Advanced Light Source (ALS)
 and R-matrix calculations \citep{Scully2005}.

In the case of Be-like and Li-like atomic oxygen systems, 
state-of-the-art  {\it ab initio}
photoionisation cross sections, resonance energies and decay rates 
for Auger inner-shell processes, were performed by Pradhan and co-workers
 \citep{Pradhan2000,Nahar2001,Pradhan2003} using the R-matrix method.
Garcia and co-workers \citep{Garcia2005} extended this work  using the 
R-matrix optical potential method within an intermediate-coupling scheme  \citep{Burke2011}
 to take account of Auger broadening of the resonances 
 in the near {\it K}-edge region.  
Photoionisation from the ground state, 
along the oxygen iso-nuclear sequence was investigated, 
in the photon energy region of the {\it K}-edge for 
both Be-like and Li-like atomic oxygen ions.
 
In the present study we focus our attention on obtaining detailed spectra for
 Be-like [O$^{4+}$ (O V)] and Li-like [O$^{5+}$ (O VI)] atomic oxygen ions 
in the vicinity of the {\it K}-edge. The current work is the completion of photoionisation cross section 
measurements and theoretical studies along the atomic oxygen iso-nuclear sequence.  
Our previous  studies on this sequence, focused on obtaining photoionisation 
cross sections for the O$^+$ and O$^{2+}$ ions \citep{Bizau2015} 
and the O$^{3+}$ ion \citep{Soleil2014}, where differences of 0.5 eV 
in the positions of the $K_{\alpha}$ resonance  lines with prior satellite observations were found.  
This will have major implications for astrophysical modelling.
\section{Experiment}\label{secExpt}

\subsection{Ion production}
The measurements were made using the MAIA (Multi-Analysis Ion Apparatus) merged-beam set-up on the PLEIADES 
photon beam line at SOLEIL, the French synchrotron radiation facility. The set-up and the experimental procedure 
have been described previously in detail \citep{Bizau2016}, and we will only give the characteristics of relevance 
for the present experiment. The oxygen ion beams are produced in a permanent magnet Electron Cyclotron Resonance Ion Source (ECRIS). 
Oxygen gas introduced in the ECRIS chamber is heated by a 12.6 GHz radio wave at a power of approximately 25W. 
A constant 4 kV bias is applied on the source to extract the ions. They are selected in mass/charge ratio by a 
dipole magnet before being collimated and merged with the photon beam in the 60 cm long interaction region. 
A second dipole magnet analyses the charge state of the ions after interaction with the photons. 
The parent ions are collected in a Faraday cup while the photo-ions (the ions which have increased 
in charge state by one in the interaction) are counted using channel plates. 
 
\subsection{Excitation source}
The photon beam is the synchrotron radiation emitted by the Apple II undulator of the PLEIADES beam line. 
The light is monochromatised using a 600 l/mm high flux grating. In the energy range considered in this work, 
the spectral purity is achieved by the cut-off of the mirrors transmission and the use of a varied groove depth for the grating. 
The photon energy is determined using an ionisation chamber \citep{Samson1967}. For this work, we used for calibration 
purpose an energy of 539.17 $\pm$ 0.15 eV for the O$_2$  $\rm 1s \rightarrow 3s \sigma$   transition as determined in \citet{Bizau2015}. 
The photon energy is corrected for the Doppler shift produced by the two counter-propagating beams. 

\subsection{Experimental procedure}
The absolute photoionisation cross sections $\sigma$ are obtained at a given photon energy using 
the procedure previously described in \citet{Bizau2016} from the relation: 
 \begin{equation}
\sigma = \frac{S e^{2}\eta {\nu } q}{I J \epsilon \int^{L}_{0} \frac{dz}{\Delta x\Delta y F(z)}},
\label{eqn1}
\end{equation}
%
%
%+++++++++++++++++++++++++++++++++++++++++++++++++++++++++++++++++++++++++++++++++++++++++++++++++++++++++++++++
%
%    Table 1 here
%
%    Here is an example of the general form of a table:
%    Fill in the caption in the braces of the \caption{} command. Put the label
%    that you will use with \ref{} command in the braces of the \label{} command.
%    Insert the column specifiers (l, r, c, d, etc.) in the empty braces of the
%    \begin{tabular}{} command.
%
%+++++++++++++++++++++++++++++++++++++++++++++++++++++++++++++++++++++++++++++++++++++++++++++++++++++++++++++++
%
 \begin{table}
\caption{\label{expt} Typical values for the experimental parameters involved 
		                 in evaluating the absolute O$^{4+}$ cross section measured 
			         at a photon energy of 554.25 eV.}
%\begin{indented}
\begin{tabular}{@{}*{7}{l}}
\hline\hline
{\it S}		&540 Hz				\\
Noise		&150 Hz				\\
$\nu$		&5.4 10$^{5}$ ms$^{-1}$	\\
Photon flux	&5.1 10$^{12}$ s$^{-1}$	\\
{\it J}			&50 nA				\\
$\epsilon$		&0.92				\\
{\it F}$_{xy}$	&5.9	10$^{4}$ m$^{-1}$	\\
\hline\hline
\end{tabular}
%\end{indented}
\end{table}

\noindent
The photo-ions counting rate $S$ is corrected from the spurious ion signal produced by 
collisional-ionisation processes using a chopper placed at the exit of the photon beam line. 
In equation (1), $\nu$ is the velocity in the interaction region of the target ions of charge state $q$, 
$I$ is the current produced by the photons on a SXUV300 IRD photodiode of efficiency $\eta$ calibrated 
at the Physikalisch-Technische Bundesanstalt (PTB) beam line at BESSY in Berlin, $J$ is the current of incident ions, 
and $\epsilon$ the efficiency of the micro-channel plates. $\Delta x \Delta y F(z)$  is an effective beam area 
(with $z$ the propagation axis of the two beams), where $F$ is the two-dimensional form factor determined 
using three sets of $xy$ slit scanners placed at each end and in the middle of the interaction region. 
The length $L$ of the interaction region is fixed by applying -1 kV bias on the tube delimiting the interaction region. 
The bias discriminates the photo-ions produced inside and outside the tube due to their different velocity. 
Typical values of the parameters involved in equation (1), measured for O$^{4+}$ target ion at a photon energy 
of 554.25 eV, are given in Table \ref{expt}. The cross sections accuracy is determined by the statistical fluctuations 
on the photo-ion and background counting rates, plus a systematic contribution resulting from the 
measurement of the different parameters in equation (1). The latter is estimated to be 15\% and is 
dominated by the uncertainty on the determination of the photon flux and the form factor. 
Two modes are used for the acquisition of the cross sections. One with no bias applied to the 
interaction tube, allowing to determine with higher statistics and spectral resolution the 
relative cross sections, which are latter normalized on the absolute cross sections 
obtained in the second mode with the voltage applied to the interaction tube.

\begin{figure*}
\begin{center}
\includegraphics[width=\textwidth]{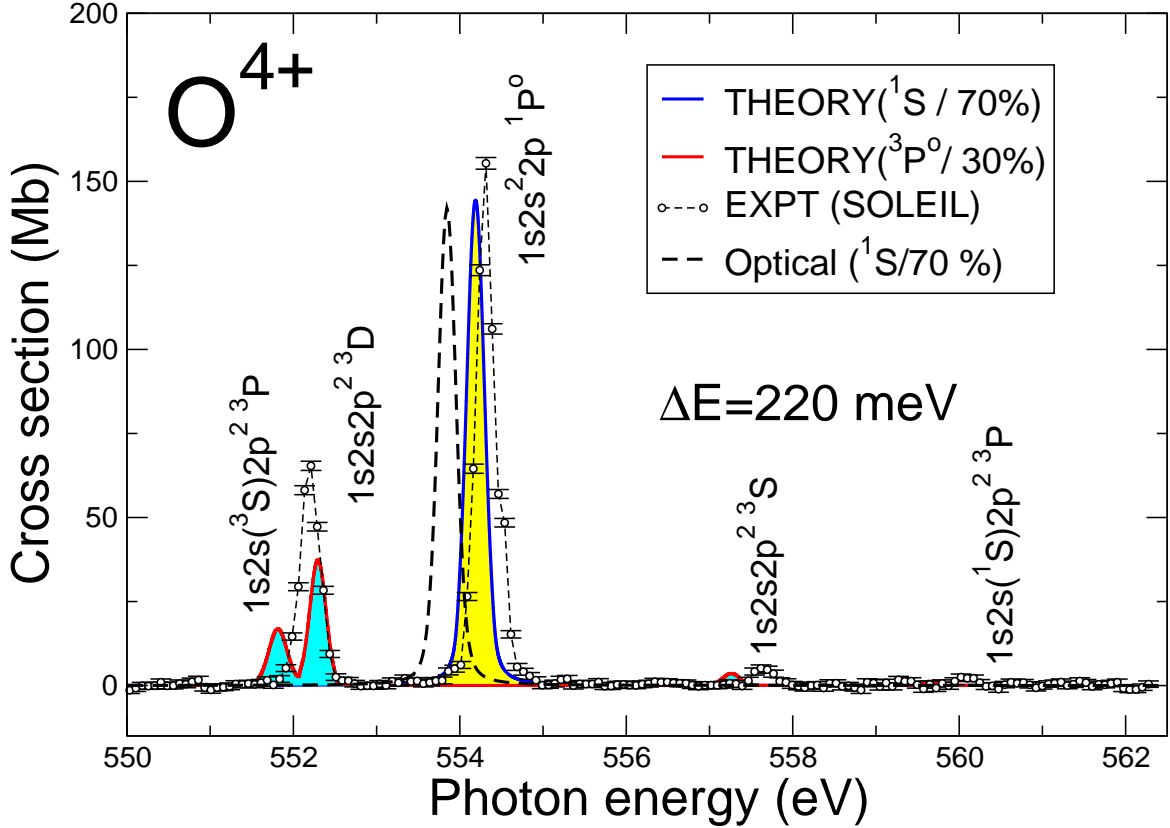}
\caption{\label{Figx1}(Colour online)  
			SOLEIL  experimental {\it K}-shell photoionisation 
			cross section of O$^{4+}$ ions 
			in the 550 - 562 eV photon energy range. 
		   	Measurements were made with a 220 meV 
		    	band-pass at FWHM. Solid points are experiment: 
		   	 the error bars represent the statistical uncertainty.
			Solid lines are the R-matrix plus pseudo-states (RMPS) 
			calculations for the ground (solid blue line) 
			and metastable state (solid red line) 
			with an appropriate admixture, see text for details.
			Dashed line (black) 
			is the intermediate coupling R-matrix calculations  
			 \citep{Garcia2005}, for the ground state only,
			using the optical potential method. 
			The strong $\rm 1s \rightarrow 2p$ resonances 
			are clearly visible in the spectra. The resonance parameters 
			are presented in Table \ref{reson1} and compared with 
			previous work in the literature. }
\end{center}
\end{figure*}

\begin{figure*}
\begin{center}
\includegraphics[width=\textwidth]{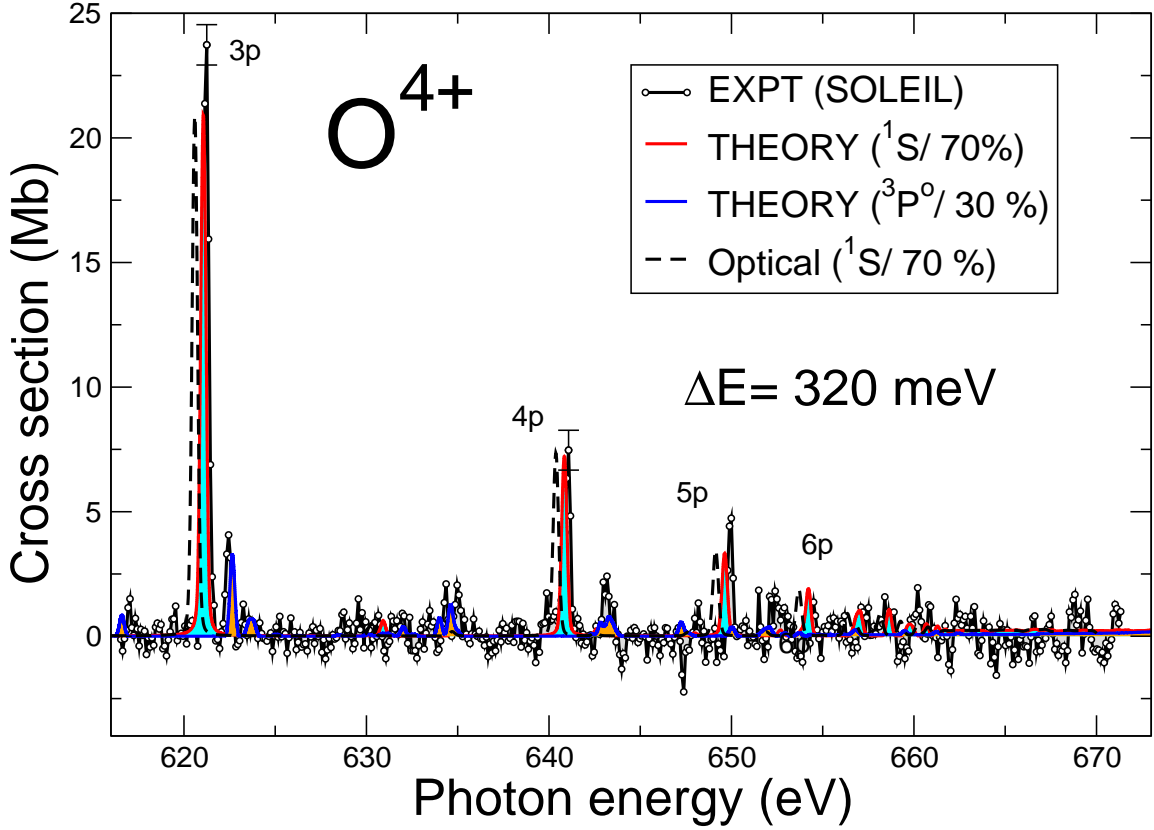}
\caption{\label{Figx2}(Colour online)  
			SOLEIL  experimental {\it K}-shell photoionisation 
			cross section of O$^{4+}$ ions in 
			the 615 - 672 eV photon energy range. 
		    	Measurements were made with a 320 meV band-pass 
		   	at FWHM. Open, solid points are experiment: 
		    	the error bars represent the statistical uncertainty.
			The solid lines are the R-matrix plus pseudo-states (RMPS) 
			calculations with an appropriate admixture,
			see text for details. Dashed line (black) 
			is the intermediate coupling R-matrix calculations  
			\citep{Garcia2005}, for the ground state only,
			 using the optical potential method.  
			The strong $\rm 1s \rightarrow np$ resonances 
			associated with the $\rm 1s^22s^2\; ^1S$ ground state 
			are clearly visible in the spectra.  The weaker features in the
			spectrum are from the $\rm 1s^22s2p\; ^3P^o$ metastable state. 
			The resonance parameters 
			are tablulated in Table \ref{reson2}. }
\end{center}
\end{figure*}

\begin{figure*}
\begin{center}
\includegraphics[width=\textwidth]{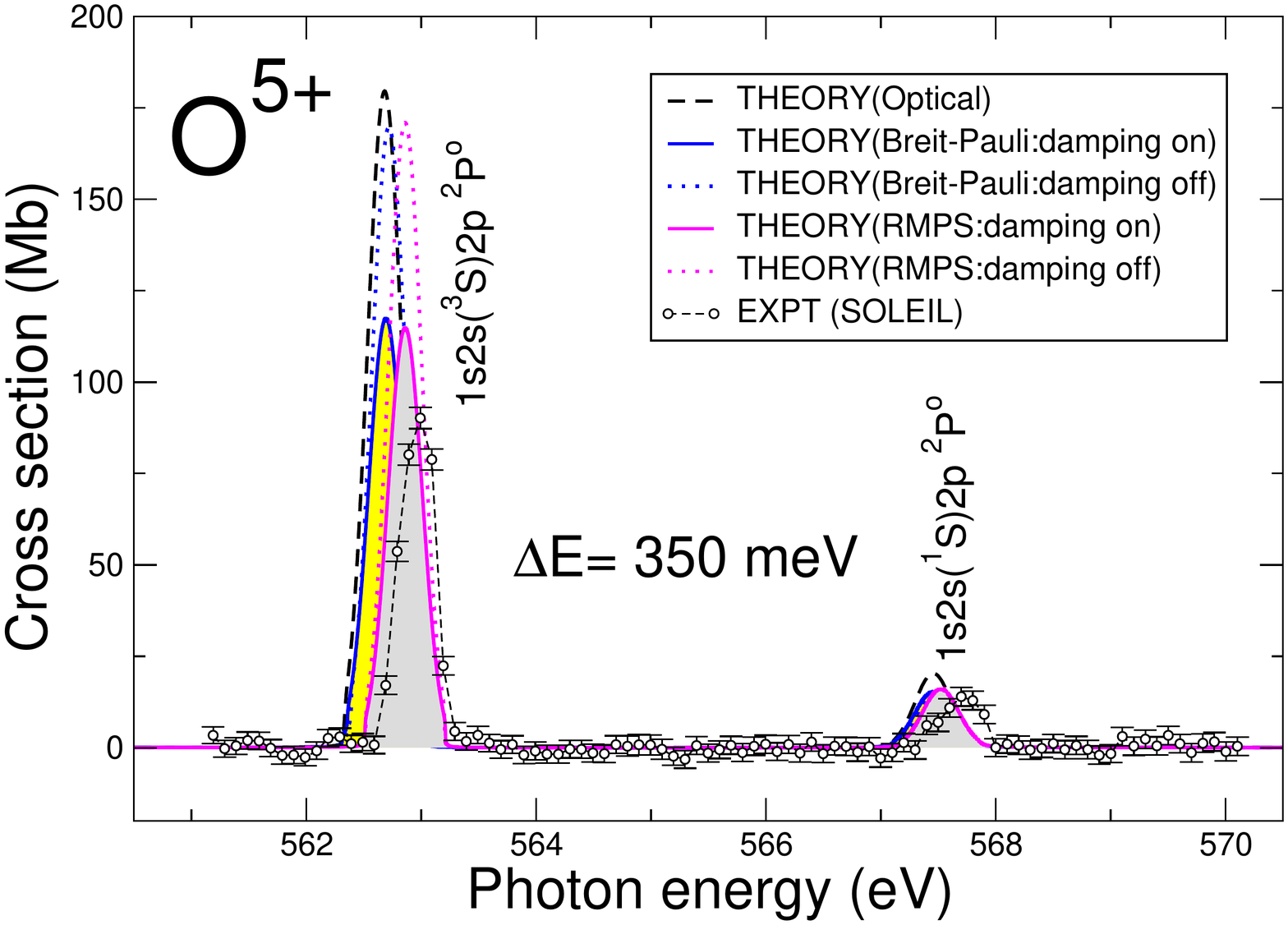}
\caption{\label{Figx3}(Colour online)  
             SOLEIL  experimental {\it K}-shell photoionisation 
			cross section of O$^{5+}$ ions in the 
			561 - 570 eV photon energy range. 
		    	Measurements were made with a 350 meV band-pass. 
		    	Solid points: the error bars give 
			the statistical uncertainty. The dashed (black) line
			are the intermediate coupling (Breit-Pauli) calculations 
			using the optical potential method \citep{Garcia2005}.
			The solid (magenta) line are the present
			 RMPS results, and the solid (blue) line 
			are those from the Breit-Pauli approximation, both of 
			which include radiation damping \citep{damp}.  
			The dotted (magneta) line and the dotted (blue) line 
			are the RMPS and Breit-Pauli approximations 
			without radiation damping.
			The strong $\rm 1s \rightarrow 2p$ resonances 
			are clearly visible in the spectra.  All the resonance 
			parameters are given in Table \ref{reson3} and compared 
			with previous work from the literature.}
\end{center}
\end{figure*}

\section{Theory}\label{secTheory}

\subsection{SCUNC: Li-like and Be-like nitrogen}\label{subsec:SU_Theory}
Previously we have used the  Screening Constant by Unit Nuclear Charge (SCUNC) method in our 
theoretical work for O$^{3+}$ \citep{Soleil2014} to provide energies and Auger decay rates 
to complement our detailed {\it ab initio} calculations. The basic equations used to 
determine the resonance energy positions and the Auger widths are summarised here 
for completeness.

In the SCUNC formalism \citep{Sakho2011,Sakho2012,Sakho2013}, one has the 
 total energy of  the core-excited states  given by,
\begin{equation}
E \left( N\ell n\ell^{\prime}; ^{2S+1}L^{\pi} \right) 
                  = -Z^2  \left[ \frac{1}{N^2} + \frac{1}{n^2}  \left( 1 - \beta  \right)^2  \right]   
\end{equation}
where $E \left( N\ell n\ell^{\prime}; ^{2S+1}L^{\pi}  \right)$ is in Rydberg units. 
In this equation, the principal quantum numbers $N$ and $n$ are respectively for the inner and the
outer electron of the He-like iso-electronic series. The $\beta$-parameters are screening constants by
unit nuclear charge expanded in inverse powers of $Z$ and are given by the expression,
\begin{equation}
 \beta \left( N\ell n\ell^{\prime}; ^{2S+1}L^{\pi}  \right)  = \sum_{k=1}^{q} f_k \left( \frac{1}{Z} \right)^k 
\end{equation}
where $f_k \left( N\ell n\ell^{\prime}; ^{2S+1}L^{\pi}  \right)$ are 
parameters to be evaluated empirically from previous experimental measurements.  
Similarly, one may get the Auger widths $\Gamma$ in Rydbergs (1 Rydberg = 13.605698 eV)  from the formula

\begin{equation}
\Gamma ({\rm Ry})  = Z^2  \left[1 
                                  -\frac{f_1}{Z} \left(   \frac{Z}{Z_0} 
                                  -\frac{1}{Z^2} \frac{(Z - Z_0)}{Z^2_0} 
                                  -\frac{1}{Z^3}  \frac{(Z - Z_0)}{Z^3_0}  \right)  \right] ^2 
\end{equation}
The measurements of M\"{u}ller and co-workers on 
Be-like boron and carbon, Li-like boron and carbon \citep{Scully2005,Mueller2009,Mueller2010,Mueller2014b} 
were used to determine all the appropriate empirical parameters 
used in the present work.

\subsection{R-matrix with pseudo-states}
The $R$-matrix with pseudo-states (RMPS) 
method \citep{mit99,Burke2011}  was used to calculate the inner-shell 
photoionisation cross sections for the atomic oxygen ions, O$^{4+}$ and O$^{5+}$, 
in their respective {\it K}-shell energy regions. 
The $R$-matrix with pseudo-sates 
method \citep{mit99,Burke2011}, using an efficient parallel implementation 
of the codes \citep{Ballance2006,McLaughlin2015a,McLaughlin2015b,McLaughlin2016a,
McLaughlin2017}
 was used to determine all the cross sections presented here.
 
 Two different basis sets and scattering models 
 were only used in our investigations on the O$^{5+}$ ion. 
 In the first collision model we used Slater orbitals where the
$\rm n$=3 physical and $\overline{\rm n}$=4 pseudo-orbitals were used for the 
residual atomic oxygen ion, O$^{6+}$, in the {\it K}-shell energy region.
This basis we designate as basis set A. 
All the Slater orbitals were generated using the 
CIV3 structure code \citep{Hibbert1975}.

In the case of O$^{5+}$ ion photoionisation, 17-levels 
in $LS$-coupling and 31-levels in intermediate-coupling,
were incorporated in our close-coupling approximation 
for the residual O$^{6+}$ ion, where semi-relativistic effects 
were included (through the Breit-Pauli approximation). 
These intermediate coupling Breit-Pauli photoionisation  
cross-section calculations were performed 
in order to gauge the influence of relativistic effects on
resonance positions, profiles, and widths for this Li-like system.

In the second approach we used the 
AUTOSTRUCTURE code \citep{Badnell1986,Badnell2011} 
to generate  the target wave functions for the subsequent photoionisation 
cross-section calculations for both of these ions. 
This we designate as basis set B.
For each of the atomic oxygen ions, O$^{4+}$ and O$^{5+}$,
physical orbitals up to $\rm n$=3  were employed. 
These were augmented with correlation type orbitals
 ${\rm \bar{n}}\ell$=4$\bar{\ell}$, $\dots$,  14$\bar{\ell}$, where 
$\ell$= 0, 1, 2, 3, and 4, i.e., $\rm s$, $\rm p$, $\rm d$, $\rm f$ and $\rm g$, 
 which are Laguerre type pseudo-states. All the computations were performed in 
$LS$-coupling.  In this approach we used 526-levels in our 
collision model for the residual O$^{5+}$ ion for photoionisation 
of O$^{4+}$. In the case of {\it K}-shell O$^{5+}$ 
photoionisation a collision model was utilised which incorporated 
120-levels of the residual O$^{6+}$ ion 
in the close-coupling approximation.
 
\clearpage
%+++++++++++++++++++++++++++++++++++++++++++++++++++++++++++++++++++++++++++++++++++++++++++++++++++++++++++++++++
%    Table  here
%
%    Here is an example of the general form of a table:
%    Fill in the caption in the braces of the \caption{} command. Put the label
%    that you will use with \ref{} command in the braces of the \label{} command.
%    Insert the column specifiers (l, r, c, d, etc.) in the empty braces of the
%    \begin{tabular}{} command.
%
%+++++++++++++++++++++++++++++++++++++++++++++++++++++++++++++++++++++++++++++++++++++++++++++++++++++++++++++++++
%
\begin{table*}
\footnotesize\rm 
\caption{\label{reson1} Be-like atomic oxygen ions, $\rm 1s \rightarrow 2p$ excitation from the 
			            $\rm 1s^22s^2[^1S] \rightarrow 1s2s^22p [^1P^{o}]$   and 
                                     $\rm 1s^22s2p[^3P^{\circ}] \rightarrow 1s2s2p^2 [^3D, ^3P, ^3S]$ core-excited states. 
			            Comparison of experiment, satellite observations and theoretical  resonance energies $E_{\rm ph}^{\rm (res)}$ (eV),
            	                     natural line widths $\Gamma$ (meV) and resonance strengths $\overline{\sigma}^{\rm PI}$ (Mb eV),
           	                     for the photo-excited n=2 states of the O$^{4+}$ ion, in the photon energy region 
        	                	             550 -- 562 eV. The SOLEIL experimental determination 
		                     of the resonances energy gives the uncertainty relative to line 2.  For the total uncertainity add 0.15 eV.   
		                     For the conversion of the satellite wavelength observations to
                                      the present energy scale $hc$ = 1239.841 974 eV nm and the absolute error $\Delta {\rm{E}} = hc \Delta \lambda / \lambda^2$ in eV were used.} 
\begin{tabular}{ccr@{\,}c@{\,}lllccccc}
\hline\hline
 Resonance    							& & \multicolumn{2}{c}{SOLEIL}            & \multicolumn{2}{c}{SATELLITE}		& \multicolumn{2}{c}{R-matrix} 	& \multicolumn{2}{c}{MCDF} 	& \multicolumn{2}{c}{OTHERS}\\
  (Label)            							& & \multicolumn{2}{c}{(Other Experiments)}	& \multicolumn{2}{c}{(Observation)} 	& \multicolumn{2}{c}{(Theory)}	& \multicolumn{2}{c}{(Theory)}	& \multicolumn{2}{c}{(Theory)}\\
 \hline
 \\
 $\rm 1s^22s2p [^3P^{\circ}] \rightarrow  1s2s[^3S]2p^2\, ^3P$				
                                  						& $E_{\rm ph}^{\rm (res)}$&& --  				& & 550.063 ${^{+0.46}_{-0.46}}^{m}$ &&551.816$^{a}$ &	& 550.200$^{d}$ 	& 551.875$^{r}$  	&\\
 		 	        						 	&           				&& 					&&                                     && 		 	    &	& 550.172$^{d}$	& 552.884$^{s}$  	&\\
~ \mycirc{0}								&					&&					&& 					&& 			&	&	 			&   551.995$^{t}$		& \\
			  							& $\Gamma$ 		 	&& --					& &  					&& 73$^{a}$ 	&	&23$^{d}$      		&  13$^{t}$ 			&\\
 	    									&           				&&  					&& 					&&  		 	&	& 	 			&   			&\\ 
										& $\overline{\sigma}^{\rm PI}$ 	&& --				&& 					&& 4.9$^{a}$	 &	&	 			&   			&\\
 \\
 $\rm 1s^22s2p  [^3P^{\circ}] \rightarrow  1s2s2p^2 \, ^3D$				
                                  						& $E_{\rm ph}^{\rm (res)}$ && 552.130 $\pm$ 0.03$^{h}$ 	& & 551.016 ${^{+0.64}_{-0.64}}^{m}$ &	&552.296$^{a}$ &	& 551.100$^{d}$  	& 552.736$^{r}$ 	&\\
 		 	        					 	        &           				&& 552.030 $\pm$ 0.80$^{k}$&& 				&&			&	& 551.175$^{d}$& 553.131$^{s}$   	&\\
~ \mycirc{1}								&					&&					&& 					&&			&	& 	 		 & 552.432$^{t}$   	&\\
									  					& $\Gamma$ 			&& 44 $\pm$ 6$^{h}$			& &  					&&42$^{a}$ 	&	&  62$^{d}$   	 &   59$^{t}$		&\\
 	    									&           				&&  					&& 					&&  		 	&	& 	 		&   			&\\ 
										& $\overline{\sigma}^{\rm PI}$ 	&& 19.8 $\pm$ 3$^{h}$	&&  					&& 9.2$^{a}$	&	&	 		&   			&\\
 \\
 $\rm 1s^22s^2  [^1S] \rightarrow  1s2s^22p \, ^1P^{\circ}$				
                                  						& $E_{\rm ph}^{\rm (res)}$&& 554.250 $\pm$ 0.00$^{h}$  & & 554.079 ${^{+0.42}_{-0.42}}^{m}$  &&554.189$^{a}$ 	&	& 553.150$^{d}$  	& 552.287$^{e}$ 	&\\
 		 	        					 		&           				&& 554.243 $\pm$ 0.24$^{i}$	&& 553.996  ${^{+0.25}_{-0.25}}^{n}$ 	&&554.550$^{b}$ 	&	& 553.241$^{d}$  	& 553.071$^{f}$  	&\\
~ \mycirc{2}								&					&& 554.144 $\pm$ 0.07$^{j}$	&& 					&&554.739$^{c'}$			&	& 553.117$^{d}$ 	& 553.971$^{r}$  	&\\
										&					&& 554.370 $\pm$ 0.20$^{l}$	&& 					&&553.853$^{g}$			&	&            		& 554.243$^{s}$   	&\\
										&					&&						&&					&&			&	&			& 554.796$^{t}$	&\\
												 	        					 		&           				&& 					&&  					&&  			&	&	 		& 554.288$^{u}$  &\\

			  							& $\Gamma$ 			&& 72 $\pm$ 4$^{h}$			&&  					&&66$^{a}$ 	&	&58$^{d}$    	&	72$^{e}$    	&\\
 										&					&&					&& 					&&63$^{c}$      	&        &				& 62$^{f}$ 	&\\
										&					&&					&&					&&66$^{g}$			&	&				& 84$^{t}$	&\\
										\\
	& $\overline{\sigma}^{\rm PI}$ && 49 $\pm$ 7$^{h}$		&& 					&& 43$^{a}$	&	&	 			&   			&\\
							&							&&					&&
	&& 44$^{g}$				&	&				&			&\\	
 \\
 $\rm 1s^22s2p  [^3P^{\circ}] \rightarrow  1s2s2p^2 \, ^3S$				
                                  						& $E_{\rm ph}^{\rm (res)}$&& 557.690 $\pm$ 0.05$^{h}$ &&556.757 ${^{+0.53}_{-0.53}}^{m}$	&&557.262$^{a}$&& 556.500$^{d}$& 557.508$^{r}$ 	&\\
 		 	        					 		&           				&& 					&&556.330 ${^{+0.10}_{-0.10}}^{o}$	&&	 		   && 556.494$^{d}$ 	 & 558.563$^{s}$  	&\\
										&					&&					&&556.050 ${^{+0.70}_{-0.70}}^{p}$  &&                       &&                           &                              &\\
~ \mycirc{3}								&					&&					&&556.732 ${^{+0.50}_{-0.50}}^{q}$ 	&&			  &&                           & 557.798$^{t}$  	&\\
			  							& $\Gamma$ 			&&  --				&&  					&&40$^{a}$ 		&	&  31$^{d}$   		&   			&\\
 	    									&           				&&  					&& 					&&  		 		&	& 	 			&   			&\\ 
										& $\overline{\sigma}^{\rm PI}$ && 2.3 $\pm$ 0.3$^{h}$	&&  					&&0.98 $^{a}$		&	&	 			&   			&\\
 \\
 $\rm 1s^22s2p [^3P^{\circ}] \rightarrow  1s2s[^1S]2p^2\, ^3P$				
                                  						& $E_{\rm ph}^{\rm (res)}$ && 560.060 $\pm$ 0.07$^{h}$  	&& 	 		&& 559.657$^{a}$ 	&	& 560.020$^{d}$ 	& 560.203$^{r}$  	&\\
 		 	        						 	&           				&& 					&& 					&& 		 		&	& 560.013$^{d}$	& 561.243$^{s}$   	&\\
~ \mycirc{4}								&					&&					&& 					&& 				&	&	 			&   			& \\
			  							& $\Gamma$ 			&& --					&&  					&& 74$^{a}$ 		&	&66$^{d}$     		&   			&\\
 	    									&           				&&  					&& 					&&  		 		&	& 	 			&   			&\\ 
										& $\overline{\sigma}^{\rm PI}$&& 0.6 $\pm$ 0.1$^{h}$		&& 					&& 0.3$^{a}$		&	&	 			&   			&\\
\hline\hline
\end{tabular}
\begin{flushleft}
$^a$R-matrix with pseudo-states (RMPS), 526-levels,basis B\\
$^b$R-matrix \citep{Berrington1997}
$^c$R-matrix \citep{Petrini1991},$^{c'}$R-matrix \citep{Pradhan2003}\\
$^d$Multi-configuration Dirac Fock (MCDF) \citep{Chen1985,Chen1986,CC1987}\\
$^e$Complex-scaled multi-reference configuration interation (CMR-CI) \citep{Yeager2012a}\\
$^f$Saddle Point + Complex Rotation (SPCR) \citep{Lin2001}, 
$^{g}$R-matrix optical potential \citep{Garcia2005}\\
$^{h}$SOLEIL present work,
$^{i}$EBIT \citep{Gu2005}, $^{j}$EBIT \citep{Schmidt2004}\\
$^{k}$Electron impact ionisation \citep{Hoffmann1990},                            
$^{l}$AUGER spectroscopy \citep{Bruch1987}\\
$^{m}${\it Chandra} observations in Mrk 279 \citep{Kaastra2004}\\
$^{n}${\it Chandra} observations in NGC 5548, Kaastra 2003 private communication \citep{Schmidt2004}\\
$^{o}${\it Chandra} observations \citep{Liao2013}, $^{p}${\it Chandra} observations \citep{Mendoza2012}\\
$^{q}${\it XMM - Newton} observations \citep{Blustin2002}\\
$^{r}$Cowan code \citep{Kaastra2004}, $^{s}$HULLAC code \citep{Kaastra2004}\\
$^{t}$SCUNC present work see text for details,
$^{u}$FAC code Gu (2010) private communication \citep{Liao2013}.\\
\end{flushleft}
\end{table*}
\clearpage

%+++++++++++++++++++++++++++++++++++++++++++++++++++++++++++++++++++++++++++++++++++++++++++++++++++++++++++++++++
%
%    Table here
%
%    Here is an example of the general form of a table:
%    Fill in the caption in the braces of the \caption{} command. Put the label
%    that you will use with \ref{} command in the braces of the \label{} command.
%    Insert the column specifiers (l, r, c, d, etc.) in the empty braces of the
%    \begin{tabular}{} command.
%
%+++++++++++++++++++++++++++++++++++++++++++++++++++++++++++++++++++++++++++++++++++++++++++++++++++++++++++++++++
\begin{table*}
\footnotesize\rm 
\caption{\label{reson2} Be-like atomic oxygen ions, $\rm 1s \rightarrow np$ excitation from the 
			            $\rm 1s^22s^2[^1S] \rightarrow 1s2s^2np [^1P^{o}]$   and 
                                     $\rm 1s^22s2p[^3P^{\circ}] \rightarrow 1s2s2pnp [^3D, ^3P, ^3S]$ ($\rm n > 2)$ core-excited states. 
		                     Comparison of experiment, satellite observations and theoretical  resonance energies $E_{\rm ph}^{\rm (res)}$ (eV),
            	                     natural line widths $\Gamma$ (meV) and resonance strengths $\overline{\sigma}^{\rm PI}$ (Mb eV),
           	                     for the photo-excited states of the O$^{4+}$ ion, in the photon energy region 
           	                     615 --  675 eV.  The SOLEIL experimental determination 
		                     of the resonances energy gives the uncertainty relative to line 2. For the total uncertainity add 0.15 eV. 
For the conversion of the satellite wavelength observations to
                                      the present energy scale $hc$ = 1239.841 974 eV nm and the absolute error  $\Delta {\rm{E}} = hc \Delta \lambda / \lambda^2$ in eV were used.}		              
\begin{tabular}{ccr@{\,}c@{\,}lllccccc}
\hline\hline
 Resonance    							& & \multicolumn{2}{c}{SOLEIL}            & \multicolumn{2}{c}{SATELLITE}		& \multicolumn{2}{c}{R-matrix} 	& \multicolumn{2}{c}{MCDF} 	& \multicolumn{2}{c}{OTHERS}\\
  (Label)            							& & \multicolumn{2}{c}{(Other Experiments)}	& \multicolumn{2}{c}{(Observation)} 	& \multicolumn{2}{c}{(Theory)}	& \multicolumn{2}{c}{(Theory)}	& \multicolumn{2}{c}{(Theory)}\\
 \hline
 \\
 \\
 $\rm 1s^22s^2  [^1S] \rightarrow  1s2s^23p \, ^1P^{\circ}$				
                                  						& $E_{\rm ph}^{\rm (res)}$&& 621.230 $\pm$ 0.05$^{e}$   & & 		--	         &&621.063$^{a}$	&	& --		& 		 	& 622.137$^{f}$\\
~ \mycirc{5}								&					&& 					&& 					&&620.604$^b$			&	& 		 	& 		  	&\\
										&					&& 					&& 					&&			&	&            		& 		   	&\\
			  							& $\Gamma$ 			&& 		--			&&  					&&80$^a$		&	&		    	&		    	&94$^{f}$\\
										&					&&					&&					&&54$^b$	&	&			&			&\\
										\\
										& $\overline{\sigma}^{\rm PI}$   && 	8.6 $\pm$ 1.3$^e$		&& 			&&8.4$^a$ 	&	&	 		&   			&\\
		&	&& 	&& 		&&8.3$^b$	&	&            		& 		   	&\\
\\
 Resonance 			
                                  						& $E_{\rm ph}^{\rm (res)}$ && 622.440 $\pm$ 0.07$^{e}$   & & 		--	 	&&622.653$^{a}$	&	& --		& 		 	&\\
~ \mycirc{6}								&					&& 					&& 					&&			&	& 		 	& 		  	&\\
										&					&& 					&& 					&&			&	&            		& 		   	&\\
			  							& $\Gamma$ 			&& 		--			&&  					&&11$^a$	 	&	&		    	&		    	&\\
										&					&&					&&					&			&	&			&			&\\
										& $\overline{\sigma}^{\rm PI}$   && 	1.4 $\pm$ 0.2$^e$&& 				&&1.1$^a$ 	&	&	 		&   			&\\
\\
$\rm 1s^22s^2  [^1S] \rightarrow  1s2s^24p \, ^1P^{\circ}$				
                                  						& $E_{\rm ph}^{\rm (res)}$&& 641.040 $\pm$ 0.06$^{e}$ & & 641.408 ${^{+0.55}_{-0.55}}^{d}$	&&640.841$^{a}$&	& --  &		 	&641.852$^{f}$\\
~ \mycirc{7}								&					&& 					&& 					&&640.388$^b$			&	& 		 	& 		  	&\\
										&					&& 					&& 					&&			&	&            		& 		   	&\\
			  							& $\Gamma$ 			&& 		--			&&  					&&78$^a$		&	&		    	&		    	&67$^f$\\
										&					&&					&&					&&52$^b$			&	&			&			&\\
										\\
										& $\overline{\sigma}^{\rm PI}$   && 	3.6 $\pm$ 0.5$^e$&& 				&&2.9$^a$ 	&	&	 		&   			&\\
\\
		&	&& 	&& 		&&3$^b$	&	&            		& 		   	&\\
\\
 Resonance 			
                                  						& $E_{\rm ph}^{\rm (res)}$&& 643.120 $\pm$ 0.09$^{e}$   & & 		--	        &&643.273$^{a}$	&	&-- 		 & 		 	&--\\
~ \mycirc{8}								&					&& 					&& 					&&			&	& 		 	& 		  	&\\
										&					&& 					&& 					&&			&	&            		& 		   	&\\
			  							& $\Gamma$ 			&& 		--			&&  					&&7$^a$		&	&		    	&		    	&\\
										&					&&					&&					&&			&	&			&			&\\
										\\
										& $\overline{\sigma}^{\rm PI}$   && 	1.2 $\pm$ 0.2$^e$&& 				&&0.2$^a$ 	&	&	 		&   			&\\
\\
 $\rm 1s^22s^2  [^1S] \rightarrow  1s2s^25p \, ^1P^{\circ}$				
                                  						& $E_{\rm ph}^{\rm (res)}$&& 649.940 $\pm$ 0.07$^{e}$ & &  650.494 ${^{+0.56}_{-0.56}}^{d}$ &&649.630$^{a}$&	&-- 	& 		 	&650.978$^{f}$\\
 ~ \mycirc{9}								&					&& 					&& 					&&649.143$^b$			&	& 		 	& 		  	&\\
										&					&& 					&& 					&&			&	&            		& 		   	&\\
			  							& $\Gamma$ 			&& 		--			&&  					&&78$^a$		&	&		    	&		    	&48$^f$\\
 										&					&&					&&					&&55$^b$		&	&			&			&\\
 										\\
& $\overline{\sigma}^{\rm PI}$   && 	2.7 $\pm$ 0.4$^d$&& 				&&1.4$^a$ 	&	&	 		&   			&\\
		&	&& 	&& 		&&1.4$^b$	&	&            		& 		   	&\\
\\
\hline\hline
\end{tabular}
\begin{flushleft}
$^{a}$R-matrix with pseudo-states (RMPS), 526-levels, basis B\\
$^{b}$R-matrix optical potential \citep{Garcia2005}\\
$^{c}$Multi-configuration Dirac Fock (MCDF) \citep{Chen1985,Chen1986,CC1987}\\
$^{d}${\it XMM - Newton} observations in NGC 3783 \citep{Behar2003}\\
$^{e}$SOLEIL present work\\
$^{f}$SCUNC present work, see text for details\\
\end{flushleft}
\end{table*}

\clearpage
%+++++++++++++++++++++++++++++++++++++++++++++++++++++++++++++++++++++++++++++++++++++++++++++++++++++++++++++++++
%
%    Table  here
%
%    Here is an example of the general form of a table:
%    Fill in the caption in the braces of the \caption{} command. Put the label
%    that you will use with \ref{} command in the braces of the \label{} command.
%    Insert the column specifiers (l, r, c, d, etc.) in the empty braces of the
%    \begin{tabular}{} command.
%
%+++++++++++++++++++++++++++++++++++++++++++++++++++++++++++++++++++++++++++++++++++++++++++++++++++++++++++++++++
%

\begin{table*}
\footnotesize\rm 
\caption{\label{reson3} Li-like atomic oxygen ions, $\rm 1s \rightarrow 2p$ excitation from the 
			      $\rm 1s^22s[^2S] \rightarrow 1s(2s2p\,^1P^{\circ})\,[^2P^{o}]$   and $\rm 1s(2s2p\,^3P^{\circ})\,[^2P^{\circ}]$ core-excited states. 
			      Comparison of experiment, satellite observations and theoretical results for the resonance energies $E_{\rm ph}^{\rm (res)}$ (eV),
            	              natural line widths $\Gamma$ (meV) and resonance strengths $\overline{\sigma}^{\rm PI}$ (Mb eV),
           	              for the photo-excited n=2 states of the O$^{5+}$ ion, in the photon energy region 
           		      560 -- 570 eV with previous investigations. 
           		      The uncertainity of the SOLEIL experiment energy of the lines is relative
           		      to line A.  For the total uncertainity, add 0.15 eV for energy calibration.
           		      For the conversion of the satellite wavelength observations to
                                      the present energy scale $hc$ = 1239.841 974 eV nm and the absolute error  $\Delta {\rm{E}} = hc \Delta \lambda / \lambda^2$ in eV were used.}
\begin{tabular}{ccr@{\,}c@{\,}lllccccc}
\hline\hline
 Resonance    							& & \multicolumn{2}{c}{SOLEIL}             & \multicolumn{2}{c}{SATELLITE}	& \multicolumn{2}{c}{R-matrix} 	& \multicolumn{2}{c}{MCDF} 	& \multicolumn{2}{c}{OTHERS}\\
  (Label)            							& & \multicolumn{2}{c}{(Other Experiments)}	& \multicolumn{2}{c}{(Observation)} 	& \multicolumn{2}{c}{(Theory)}	& \multicolumn{2}{c}{(Theory)}	& \multicolumn{2}{c}{(Theory)}\\
 \hline
 \\
 $\rm 1s^22s  [^2S] \rightarrow  1s(2s2p\ ^3P^{\circ}) \, [^2P^{\circ}]$				
                                  					        & $E_{\rm ph}^{\rm (res)}$ &	& 562.940 $\pm$ 0.00$^{j}$ 	&&562.662 ${^{+0.70}_{-0.70}}^{q}$&
										&562.860$^{a}$       &      &562.018$^{d}$      	&562.829$^{e}$ &\\
 		 	     				 		&           				&	& 563.000 $\pm$ 1.00$^{k}$ 	&&562.899 ${^{+0.10}_{-0.10}}^{r}$ &&562.709$^{a'}$	&	&563.039$^{d}$        &562.415$^{f}$  &\\
 		 	        					 		&           				&	& 563.119 $\pm$ 0.20$^{n}$	&&562.950 ${^{+0.08}_{-0.08}}^{s}$ &&562.297$^{b}$	&	&562.803$^{d}$    	&563.440$^{g}$ &\\
~ \mycirc{A}								&					&	& 563.068 $\pm$ 0.04$^{o}$	&&562.828 ${^{+0.06}_{-0.05}}^{t}$  &&562.680$^c$				&	&				&563.057$^{h}$ &\\
 		 	        					 		&           			         &	& 563.053 $\pm$ 0.15$^{p}$ 	&&563.309 ${^{+0.26}_{-0.26}}^{u}$ &&	      			&	& 				&562.534$^{x}$ &\\
 		 	        					 		&           			         &	&563.100$\pm$ 0.20$^{m}$		&&562.287 ${^{+0.40}_{-0.40}}^{v}$ &&	      			&	& 				&562.705$^{i'}$  &\\
										&					&	&						&&562.465 ${^{+0.10}_{-0.10}}^{y}$ &&				&	&                             	&                         &\\
										&					&	&						&&562.912 ${^{+0.24}_{-0.23}}^{z}$ &&				&	&				&                         &\\
										&					&	&						&&563.053 ${^{+0.51}_{-0.51}}^{z'}$							&&				&	&				&                          &\\
\\
			  							& $\Gamma$ 			&	& 	--					& &  						&       & 6$^{a}$ 		&	&5$^{d}$          		& 6$^{e}$    &\\
 										&					&	&						&& 						&	& 6$^{a'}$    		&     	&				&4$^{f}$     & \\
										&					&	&						&& 						&	& 4$^{b}$      		&      	&				&6$^{g}$     & \\
										&					&	&						&&						&	& 9$^{c}$				&	&				&6$^{i'}$	  &\\
\\
										& $\overline{\sigma}^{\rm PI}$ 	&	& 	36  $\pm$ 5$^{j}$	&& 						&	&42.52$^{a}$		&	&	 			&   		&\\
										&					&	&						&&						&	& 42.52$^{a'}$           	&      &                               &                  & \\																			&					&	&&&			&	& 71.3$^{c}$           	&      &                               &                  & \\
										\\
 $\rm 1s^22s  [^2S] \rightarrow  1s(2s2p\,^1P^{\circ}) \,[ ^2P^{\circ}]$				
                                  						& $E_{\rm ph}^{\rm (res)}$&	& 567.620 $\pm$ 0.05$^{j}$ 	&& 566.915 ${^{+0.40}_{-0.40}}^{v}$ && 567.519$^{a}$ 	&	&568.499$^{d}$    	&567.420$^{e}$  &\\
 		 	        					 		&           				&	& 566.081 $\pm$ 0.12$^{l}$	&& 					 	&	& 567.445$^{a'}$ 	&	&		 		&567.420$^{f}$  &\\
~ \mycirc{B}								&					&	& 567.719 $\pm$ 2.00$^{m}$	&& 						&	& 566.987$^{b}$   	&	& 		          	&568.117$^{g}$  &\\
 										&					&	& 568.000 $\pm$ 2.00$^{n}$	&& 						&	& 567.450$^{c}$ 	&     	& 				&567.783$^{i}$   & \\
									         &					&	& 567.563 $\pm$ 0.26$^{p}$						&&						&	&                             	&	&				&567.762$^{i'}$	  &\\
\\
			  							& $\Gamma$ 			& 	& 	--					&&  						&      & 46$^{a}$ 		&	&49$^{d}$    		& 46$^{f}$  	&\\
 										&					&	&						&& 						&	& 46$^{a'}$       		&	&				& 50$^{g}$ 	&\\
 										&				        &	&						&& 					&	& 45$^{b}$  		&    	& 				& 45$^{i}$		&\\
										&		&				&				&&									&	& 48$^{c}$		&	    &				& 59$^{i'}$	&\\
\\
										& $\overline{\sigma}^{\rm PI}$ 	&&     7.4 $\pm$ 1$^{j}$ 			&& 						&	& 6.4$^{a}$		&	&	 			&   	          &\\
										&					&	&						&&						&	& 6.3$^{a'}$           	&      &                               &                  &\\
										&					&	&						&&						&	& 8.6$^{c}$           	&      &                               &                  &\\
\hline\hline
\end{tabular}
\begin{flushleft}
$^{a}$R-matrix with pseudo-states (RMPS), 120-levels, basis B, 
$^{a^{\prime}}$R-matrix, Breit-Pauli, 31-levels, basis A\\
$^{b}$R-matrix \citep{Pradhan2003}\\
$^{c}$R-matrix, optical potential method, intermediate coupling \citep{Garcia2005}\\
$^{d}$Multi-configuration Dirac Fock (MCDF) \citep{Chen1985,Chen1986, Chen1987}\\
$^{e}$Complex-scaled multi-reference configuration interation (CMR-CI) \citep{Yeager2012b}\\
$^{f}$Saddle Point + Complex Rotation (SPCR) \citep{BW2000}\\ 
$^{g}$Saddle Point + Complex Rotation (SPCR) \citep{Wu1991}\\ 
$^{h}$Intermediate-coupling \citep{Gabriel1972}\\ 
$^{i}$Saddle Point Method (SPM) \citep{Davis1989}, 
$^{i'}$SCUNC present work, see text for details\\
$^{j}$SOLEIL present work,
$^{k}$Auger Spectroscopy \citep{Bruch1979},
$^{l}$EBIT \citep{Gu2005}\\
$^{m}$Auger Spectroscopy \citep{Bruch1987},
$^{n}$Electron-impact ionisation experiments \citep{Hoffmann1990}\\
$^{o}$EBIT \citep{Schmidt2004},
$^{p}$Laser plasma experiment \citep{Tondello1977}\\
$^{q}${\it Chandra} observations \citep{Mendoza2012}\\
$^{r}${\it Chandra} observations \citep{Yao2009}, $^{s}${\it Chandra} observations \citep{Gatuzz2013a,Gatuzz2013b}\\
$^{t}${\it Chandra} observations \citep{Liao2013}\\
$^{u}${\it Chandra} observations in NGC 5548, Kaastra (2003) private communication \citep{Schmidt2004}\\
$^{v}${\it Chandra} observations in NGC 5548 \citep{Kaastra2002},
$^{x}$FAC code Gu (2010) private communication \citep{Liao2013}\\
$^{y}${\it Chandra} observations in MCG-6-30-15 \citep{Lee2001}\\
$^{z}${\it XMM-Newton} observations in Cyg X-2 \citep{cabot2013}\\
$^{z'}${\it XMM-Newton} observations in Mkn 501 RGS1 \citep{Nicastro2016b}\\
\end{flushleft}
\end{table*}
\clearpage

\subsection{Be-like atomic oxygen ions}
For the O$^{4+}$ ion we used the R-matrix with 
pseudo-states method to perform all the 
photoionisation cross sections. All the photoionisation cross section 
calculations were performed in $LS$-coupling. 
We note that the ion-beam in the SOLEIL experiments
contain both ground state and metastable states.
Photoionisation cross-section calculations were carried out
for both these initial states of O$^{4+}$;  
$\rm 1s^22s^2 \, ^1S$ ground state and 
the $\rm 1s^22s2p\, ^{3}P^{\circ}$ metastable state. 
Thirty-five continuum orbitals were 
utilised in the collision calculations.   
In our collision model we retained
526-levels in the close-coupling expansion and 
basis set B was utilised for this scattering approximation.    
A boundary radius of 13.838 Bohr 
was required to cater for the very diffuse 
pseudo-states.  In the outer region 
we used an energy mesh of 1.36 $\mu$eV 
to fully  resolve all the fine resonance 
features in the cross sections. 
%+++++++++++++++++++++++++++++++++++++++++++++++++++++++++++++++++++++++++++++
%
%    Table here
%
%    Here is an example of the general form of a table:
%    Fill in the caption in the braces of the \caption{} command. Put the label
%    that you will use with \ref{} command in the braces of the \label{} command.
%    Insert the column specifiers (l, r, c, d, etc.) in the empty braces of the
%    \begin{tabular}{} command.
%
%+++++++++++++++++++++++++++++++++++++++++++++++++++++++++++++++++++++++++++++
%
\begin{table*}
\caption{\label{osc-all} Li-like and Be-like ions, 
			integrated oscillator strength for
			the core-excited states arising from respectively the 
			ground configurations $\rm 1s^22s$ and $1s^22s^2$ of each ion. 
			The table shows a comparison of the present experimental
			and theoretical results for the integrated oscillator 
			strengths $f$ for the dominant core photo-excited n=2 states 
			of the first four ions of each sequences 
			(Z, is the atomic number of the ion)
			 along with previous investigations.}
 %\noindent
\begin{tabular}{cccrcllcl}
\hline\hline
Z &ION	  		&   				  & 	    &EXPERIMENT 	& 		&THEORY		& &  \\
 \hline\hline
Li-like &$\rm 1s^22s\,^2S \rightarrow 1s[2s2p\,^3P]\,^2P^o$				& 			&							&Resonance				& 			& &	\\
  			\\			
5 & B$^{2+}$			&$f$				&				&0.465  $\pm$ 0.09$^{a}$
 & 				&0.413$^{a'}$		& &  \\
 \\
6 & C$^{3+}$	  		&$f$   				  & 		    & 0.483 $\pm$ 0.10$^{b}$ 	& 				&0.485$^{b'}$		& &  \\
				\\ 		
7 & N$^{4+}$	  		& $f$   				  & 		         & 0.413 $\pm$ 0.07$^{c}$	& 				&0.546$^{i}$		& &  \\
                \\
8 & O$^{5+}$	  		& $f$   				  & 		         & 0.328 $\pm$ 0.05$^{d}$	& 				&0.387$^{j}$		& &  \\
\\
 Be-like &$\rm 1s^22s^2\,^1S \rightarrow 1s2s^22p\,^1P^o$				&			&							&	Resonance			& 			& &	\\
  			\\			
5 & B$^{+}$			&$f$				&				&0.641  $\pm$ 0.13$^{e}$
 & 				&0.413$^g$		& &  \\
 \\
6 & C$^{2+}$	  		&$f$   				  & 		    & 0.624 $\pm$ 0.10$^{f}$ 	& 				&0.447$^h$		& &  \\
				\\ 		
7 & N$^{3+}$	  		& $f$   				  & 		         & 0.650 $\pm$ 0.10$^{c}$	& 				&0.546$^{i}$		& &  \\
                \\
8 & O$^{4+}$	  		& $f$   				  & 		         & 0.640 $\pm$ 0.09$^{d}$	& 				&0.559$^{j}$		& &  \\

\hline\hline
\end{tabular}
\begin{flushleft}
$^{a}$ALS, experiment \citep{Mueller2010}, 
$^{b}$ALS, experiment \citep{Mueller2009}\\
$^{a'}$R-matrix, theory \citep{Mueller2010}, 
$^{b'}$R-matrix, theory \citep{Mueller2009}\\
$^{c}$SOLEIL, experiment \citep{Soleil2013}, 
$^{d}$SOLEIL, experiment present work.\\
$^{e}$ALS, experiment \citep{Mueller2014b}, 
$^{f}$ALS, experiment \citep{Scully2005}\\
$^{g}$R-matrix,theory \citep{Mueller2014b},
$^{h}$R-matrix,theory \citep{Scully2005}\\
$^{i}$R-matrix,theory \citep{Soleil2013},
$^{j}$R-matrix,theory present work
\end{flushleft}
\end{table*}
%
%+++++++++++++++++++++++++++++++++++++++++++++++++++++++++++++++++++++++++++++++++++++++++++++++++++++++++++++++++
%
%    Table  here
%
%    Here is an example of the general form of a table:
%    Fill in the caption in the braces of the \caption{} command. Put the label
%    that you will use with \ref{} command in the braces of the \label{} command.
%    Insert the column specifiers (l, r, c, d, etc.) in the empty braces of the
%    \begin{tabular}{} command.
%
%+++++++++++++++++++++++++++++++++++++++++++++++++++++++++++++++++++++++++++++++++++++++++++++++++++++++++++++++++
%
\begin{table*}
\caption{\label{compare} Comparison of the energies for the $K_{\alpha}$ 
 						resonance line (centroid in eV), ions in their ground state, 
                        along the atomic oxygen isonuclear sequence.  
                        The entries in the table are
                       from synchrotron radiation light source (SR) and 
                       EBIT measurements, compared to available 
                       Chandra and XMM-Newton satellite 
                        observations. For conversion of the satellite 
                        wavelength observations to the present 
                        energy scale $hc$ = 1239.841 974 eV nm was used.}
%\begin{indented}
\begin{tabular}{cllcllcl}
\hline\hline
Atomic Oxygen			&SR		&         &EBIT 	&	&   	&{\it Chandra and XMM}&\\ 
\hline			
Ionised state	&E (eV)		&E(eV)     & & $\Delta$E(eV)	&E(eV)		 &	&$\Delta$E(eV) \\
\hline\hline
OI   &526.79(4)$^a$	 &		&&     		        &527.39(2)$^g$		&   &-0.60\\
     &                &     &&                   &527.37(40)$^o$     &   &-0.58\\
     &				  &		&&					&527.44(18)$^p$     &   &-0.65\\
     \\   
OII	 &530.50(13)$^b$	 &		&& 	                &530.97(3)$^g$		&   &-0.47\\
     &               &      &&                  & 530.98(40)$^o$     &   &-0.48\\
     \\
OIII	 &536.86(13)$^b$	 &537.42(9)$^e$	&&-0.55	   	&537.94(2)$^g$ 	  	&   &-1.08\\
\\
OIV  &544.52(13)$^c$	 &545.20(10)$^e$	&&-0.68	 	&546.22(8)$^g$	 	&   &-1.74\\
     &               &              &&           &546.43(24)$^p$     &  &-1.91\\
\\
OV   &554.25(18)$^d$	 &554.14(7)$^f$  &&~0.11	  	&556.33(9)$^{g,\ast}$&   &-2.08\\
	&		         &			     &&  		&554.00(25)$^h$		&   &-0.25\\
    &        		 &               &&         &554.08(42)$^i$ 	    &   &~0.17\\
    \\
OVI &562.94(18)$^d$	 &563.07(4)$^d$	&&-0.13	 	&562.83(6)$^g$ 		&   &~0.11\\
			&			&		&&		&562.90(10)$^j$ 	&   &~0.04\\
			&			&		&&		&562.95(10)$^k$ 	&   &-0.01\\
			&			&		&&		&563.31(26)$^h$ 	&   &-0.63\\
			&			&		&&		&562.29(40)$^l$ 	&   &~0.65\\
			&			&		&&		&562.47(10)$^m$ 	&   &~0.47\\
			&           &       &&     	&562.90(25)$^n$ &   &~0.04\\
			&			&		&&		&563.05(51)$^p$ &   &-0.11\\
\hline\hline	
\end{tabular}		
%\end{indented}
\begin{flushleft}
$^a$Advanced Light Source \citep{Stolte2013,Oxygen2013}\\
$^b$SOLEIL \citep{Bizau2015}\\
$^c$SOLEIL \citep{Soleil2014,Bizau2015}\\
$^d$SOLEIL present results\\
$^e$EBIT \citep{Gu2005}\\
$^f$EBIT \citep{Schmidt2004}\\
$^g${\it Chandra} \citep{Liao2013}\\
$^h${\it Chandra} Kaastra 2003, private communication, \citep{Schmidt2004}\\
$^i${\it Chandra} \citep{Kaastra2004}\\
$^j${\it Chandra} \citep{Yao2009}\\
$^k${\it Chandra} \citep{Gatuzz2013a,Gatuzz2013b}\\
$^l${\it Chandra} \citep{Kaastra2002}\\
$^m${\it Chandra} \citep{Lee2001}\\
$^n${\it XMM - Newton} \citep{cabot2013}\\
$^o${\it Chandra and XMM-Newton} \citep{Nicastro2016a}\\
$^p${\it Chandra and XMM-Newton} \citep{Nicastro2016b}\\
$^{\ast}$OV line observed by \cite{Liao2013}, is the OII $K_{\beta}$ line 
(at 556.48(50) eV), confirmed recently by \cite{Nicastro2016a}.\\
\end{flushleft}
\end{table*}
\subsection{Li-like atomic oxygen ions}
For the O$^{5+}$ ion we used the 
R-matrix with pseudo-states method (RMPS)
and the Breit-Pauli approximation 
to perform all the photoionisation cross sections. 
Basis set A was used with 17 levels 
included in $LS$-coupling and 
31 levels in the intermediate-coupling 
approximation for the residual  O$^{6+}$ ion.  
In the SOLEIL measurements, 
for this ion, only the ground-state 
is present, so cross section calculations were performed 
for the $\rm 1s^22s \, ^2S$ ground state of O$^{5+}$ 
in $LS$ and in intermediate coupling 
taking into account relativistic effects through 
the Breit-Pauli (BP) approximation. 

Thirty-five continuum orbitals 
were utilised in the collision calculations and 
a boundary radius of 6.0 Bohr to accommodate 
the diffuse $\rm n$=4 orbitals for basis A. 
For the R-matrix with pseudo-states approach,  
120-levels were retained in the close-coupling approximation, 
with basis set B.
A boundary radius of 11.925 Bohr was 
required to cater for the very diffuse pseudo-states and
thirty-five continuum orbitals used for the collision calculations.   

\begin{figure*}
\begin{center}
\includegraphics[width=\textwidth]{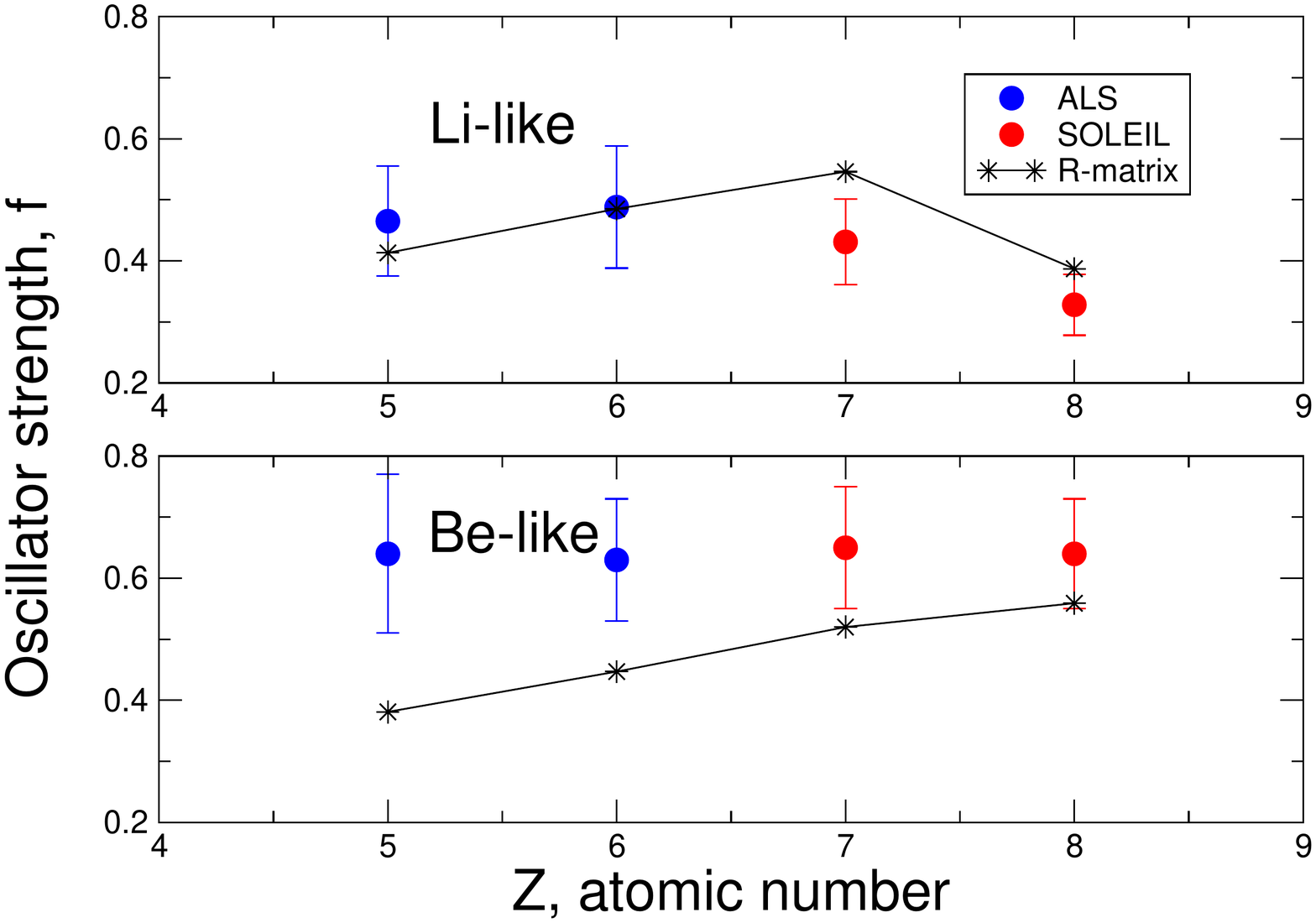}
\caption{\label{Figx4}(Colour online)  
           Comparison of the integrated oscillator strengths $f$ obtained
           from synchrotron radiations measurements with R-matrix calculations. 
           The solid circle are from the ALS (blue) and SOLEIL(red)
           radiation facilities, respectively, for the
           Li-like and Be-like iso-electronic sequences.
           The asterisk values (black) are from  R-matrix calculations.}
\end{center}
\end{figure*}
Photoionization cross-section calculations for O$^{5+}$ 
ions were performed both in $LS$ and intermediate coupling. 
The intermediate coupling calculations were carried 
out using the semi-relativistic Breit-Pauli 
approximation which allows for relativistic effects to
be included. Radiation-damping \citep{damp} effects were also
included within the confines of the R-matrix approach \citep{Burke2011}
for completeness as this affects only narrow resonances 
found in cross-sections for highly charged systems.
For the O$^{5+}$ ion an energy mesh of 680 $\mu$eV 
resolved all the resonance features in the spectra for the 
photon energy range 560 - 575 eV. 
		%`
\begin{figure*}
\begin{center}s
\includegraphics[width=\textwidth]{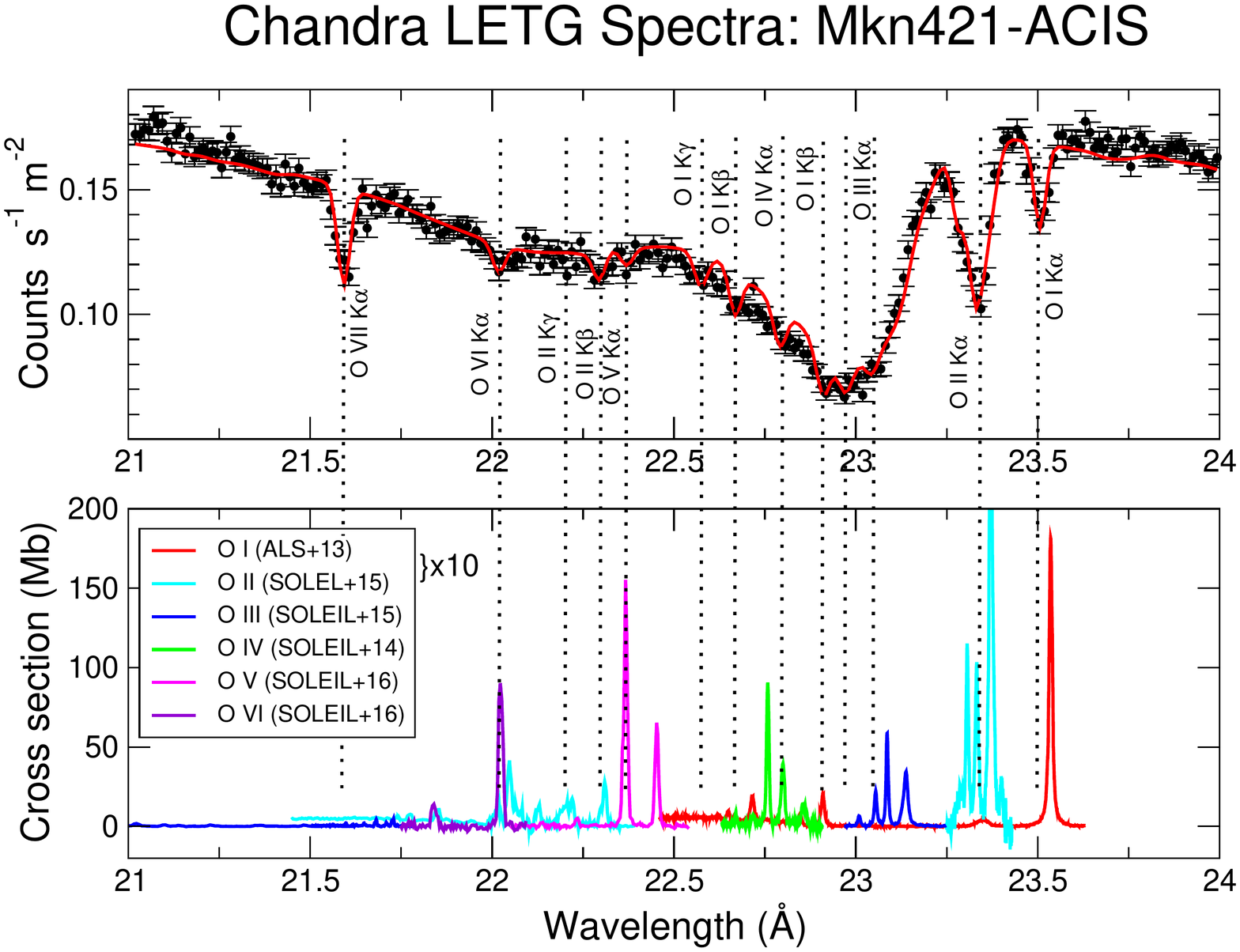}
\caption{\label{Figx5}(Colour online)  
            		Top panel, Chandra LETG spectra for the blazar Mkn 421 ACIS
		         in the wavelength 21 - 24 {\AA} composed from several different 
		         exposures with the Chandra satellite 		
		         \citep{Nicastro2016a}.
		         Bottom panel,  for the same wavelength range, 
		         measurements are illustrated from a combination 
		         of ground based experimental spectra made at
		         various synchrotron radiation facilities; 
		         Advanced Light Source (O I) \citep{Stolte2013,Oxygen2013}, 
		         SOLEIL (O II and O III) \citep{Bizau2015} 
                         SOLEIL (O IV) \citep{Soleil2014}
		         and the present SOLEIL (OV and O VI) results. The strongest lines 
                         in the synchrotron measurements are from the ground state of each 
                         ionised stage of atomic oxygen.
                         The weaker ones in the measurements are
                         from the metastable states of each ionic complex.
                         The OI and OII results are scaled by a factor of 10.}
\end{center}
\end{figure*}	
%+++++++++++++++++++++++++++++++++++++++++++++++++++++++++++++++++++++++++++++++++++++++++++++++++++++++++++++++++
%
%    Table  here
%
%    Here is an example of the general form of a table:
%    Fill in the caption in the braces of the \caption{} command. Put the label
%    that you will use with \ref{} command in the braces of the \label{} command.
%    Insert the column specifiers (l, r, c, d, etc.) in the empty braces of the
%    \begin{tabular}{} command.
%
%+++++++++++++++++++++++++++++++++++++++++++++++++++++++++++++++++++++++++++++++++++++++++++++++++++++++++++++++++
%
\begin{table}
\caption{\label{calibrate} Comparison of the calibrated energies  
			from the EELS and PES experimental methods \citep{Bizau2015}. 
			The energy difference between the 
			two experimental methods is at most 0.22 eV.} 
%\begin{indented}
\begin{tabular}{cccc}
\hline\hline
O$_2$ K-shell			&EELS		  	&PES 	 	& $\Delta$E (eV)\\ 
\hline\hline			
$\rm 1s \rightarrow \pi^{\ast}$	&530.80(20)$^a$		&530.70(15)	&-0.10			\\
\\
$\rm 1s \rightarrow 3\sigma$	&538.95(4)$^b$		&539.17(15)	&+0.22			\\ 
\hline\hline	
\end{tabular}		
%\end{indented}
\begin{flushleft}
$^a$\cite{Brion1980}\\
$^b$\cite{Tanaka2008}
\end{flushleft}
\end{table}
\section{Results and Discussion}\label{secResults}
Fig \ref{Figx1} illustrates the experimental and theoretical results 
for the O$^{4+}$ ion in the photon energy region 550 - 562 eV 
where the strong $\rm 1s \rightarrow 2p$ resonances are found.  
The theoretical cross sections were convoluted with 
an appropriate Gaussian profile having a 
similar band width to the SOLEIL 
measurements;  220 meV for the region 550 - 565 eV.  
An admixture of the ground and metastable states 
was applied to simulate experiment.  We  
found that 70\% of the $\rm 1s^22s^2\; ^1S$ ground and
30 \% of the $\rm 1s^22s2p\; ^3P^o$ metastable states 
suitably matched theory with experiment.
In Fig. \ref{Figx2} we present the 
photon energy region 615  -- 670 eV 
where the $\rm 1s \rightarrow np$ resonance 
features are located.  Here again a similar 
admixture (70\% ground state and 30\% metastable state) 
for the theoretical cross sections, after convolution 
by a Gaussian profile with 320 meV FWHM,
simulates the experimental results very well.  
Tables \ref{reson1} and \ref{reson2} 
compare our experimental and theoretical 
results for the resonance parameters for this Be-like systems 
with other theoretical, experimental and satellite observations.  
Figs \ref{Figx1} and \ref{Figx2}
includes the intermediate-coupling calculations 
of Garcia and co-workers \citep{Garcia2005} performed
using the optical potential method (dashed black line) convolved 
by the Gaussian profiles and weighted by 70\% 
for the ground state only for completeness.

The results for the resonance strengths 
$\bar{\sigma^{\rm PI}}$ found in the  O$^{4+}$
 spectra are presented in  
Tables \ref{reson1} and \ref{reson2}
and the resonance strengths have been 
weighted by the 70\% and 30\% 
admixture for the ground and metastable states
to compare directly with experiment. 

For the O$^{4+}$ ion it  should be  
noted that from a comparison of the experimental and  
R-matrix resonance strengths $\bar{\sigma^{\rm PI}}$ 
(see Fig. \ref{Figx1} and Fig. \ref{Figx2}) 
 excellent agreement, particularly 
 for the strong resonant features 
 (due to the $\rm 1s^22s^2\; ^1S$ ground state)
 in the spectrum is achieved.  It can be
 seen from  Tables \ref{reson1} and \ref{reson2}, 
the agreement between experiment and 
theory for the weaker resonant features in 
the spectra of O$^{4+}$ 
(due to the $\rm 1s^22s2p\; ^3P^o$ metastable) is not as good
 compared to the ground state. The weak resonances 6 and 8 
 in Table \ref{reson2} (originating from 
 the $\rm 1s^22s2p\; ^3P^o$ metastable state)
 may be identified with the aid of quantum defect theory.
 The weak resonance profiles of the remaining members of the Rydberg series 
 emanating from the $\rm 1s^22s2p\; ^3P^o$ metastable state make them 
 difficult to resolve experimentally.

Fig. \ref{Figx3} shows a comparison of the SOLEIL cross-section measurements 
made with a band-pass of 350 meV FWHM for the O$^{5+}$ ion 
with our theoretical results which include and exculde radiation damping \citep{damp}.  
In order to compare directly with experiment, theory has been convoluted with 
a Gaussian profile having a similar width of 350 meV at  FWHM. The results from 
both the Breit-Pauli (BP) approximation (using basis A) and those 
from the R-matrix with pseudo-states method (using basis B) are shown.
As seen from Fig. \ref{Figx3} the calculations of 
 Garcia and co-workers \citep{Garcia2005}
 performed (using the optical potential approach) in intermediate-coupling, 
 would appear to not include radiation damping \citep{damp}, as they 
 are similar to our Breit-Pauli results without radiation damping, 
 and as such overestimate the resonance 
 strength of the narrow resonance located by experiment at 562.94 eV.  
 As illustrated in Fig. \ref{Figx3} and from the results 
 presented in Table \ref{reson3}, 
the R-matrix with pseudo-states (RMPS) cross sections
(that include radiation damping) give best agreement with the
 present measurements from the SOLEIL 
 radiation facility for this ion in the photon energy region 
 where the resonances of O$^{5+}$ are located. 
 
An additional check on the theoretical data is the comparison of 
the integrated oscillator strength $f$ or the resonance strength with experiment. 
The quantity $f$ for the theoretical and experimental spectra may be 
determined for each resonance
using  \citep{Shore1967,Fano1968,berko1979},
\begin{equation}
f  = 9.1075 \times 10^{-3} \bar{\sigma^{\rm PI}}. 
\end{equation}
The resonance strength in the photoionisation 
cross-section $\bar{\sigma^{\rm PI}}$ is defined as
\begin{equation}
\bar {\sigma^{\rm PI}}= \int_{E_1}^{E_2} \sigma (h\nu) dh\nu,
\end{equation}
where [$E_1$,$E_2$] is the photon energy range 
over which each resonance profile extends.
  
Fig. \ref{Figx4} illustrates results for the  oscillator strengths $f$
along  the Li-like and Be-like iso-electronic sequences, 
for the strong $K_{\alpha}$ transitions originating 
from the ground state of each ion. Table \ref{osc-all} tabulates these integrated 
oscillator strengths $f$ results obtained from synchrotron 
radiation (SR) measurements (ALS and SOLEIL) 
and from R-matrix calculations. For the Be-like sequence the values 
have been corrected for ground state populations.  
As can be seen from Table \ref{osc-all} and 
Fig. \ref{Figx4}, apart from Be-like boron and to a lesser extent Be-like carbon, 
the agreement between theory and experiment along both 
iso-electronic sequences is quite satisfactory giving 
further confidence in our current work.

\section{Comparison with satellite observations}\label{secSatellite}
In Tables \ref{reson1}, \ref{reson2}, and \ref{reson3} comparisons 
are made with the  available experimental and satellite observations 
in the literature for resonance energies 
for Be-like and Li-like atomic oxygen ions.  
One can clearly see  various discrepancies 
between our present ground based measurements, 
made at SOLEIL, and the Chandra and XMM-Newton satellite observations. 
We note that for both atomic ions, the R-matrix with pseudostates 
results favour those from the SOLEIL radiation facility.

Fig \ref{Figx5} illustrates a 
comparison of various ground based synchrotron 
cross section measurements (ALS and SOLEIL) 
for the atomic oxygen isonuclear sequence 
in  the wavelength region 21 -- 24 {\AA} 
with the low-energy-transmission-grating (LETG) spectra
observed by Chandra for the blazar Mkn 421 ACIS \citep{Nicastro2016a}.   
One clearly sees, especially for low charged stages of oxygen ions, 
discrepancies in the resonance energies for the strong $K_{\alpha}$ 
lines in the Chandra spectra compared to the different ground based
 light source measurements which we discuss and quantify. 
The satellites observe OI to 
OIV lines (with a shift) and noshift in the OV and OVI 
lines, as seen in the comparison with the 
spectra for the blazar Mkn 421 \citep{Nicastro2016a}, 
shown in Fig \ref{Figx5}.  
Similar effects are also seen in the spectra 
of H 2356-309 LETG, H 2356-309 RGS1, and Mkn 501 RGS1, 
for the wavelength region 21 -- 24 {\AA} 
in the recent observations of Nicastro 
and co-workers \citep{Nicastro2016a,Nicastro2016b}.
We note similar differences are seen 
in the wavelength region 21 -- 24 {\AA}, for satellite observations
in the high-energy-transmission grating (HETG) spectra 
observed by Chandra for the x-ray binary XTE J1817-332 
\citep{Gatuzz2013a,Gatuzz2013b}, although 
no OIV or OV K-lines are observed.  
 
\begin{figure*}
\begin{center}
\includegraphics[width=\textwidth]{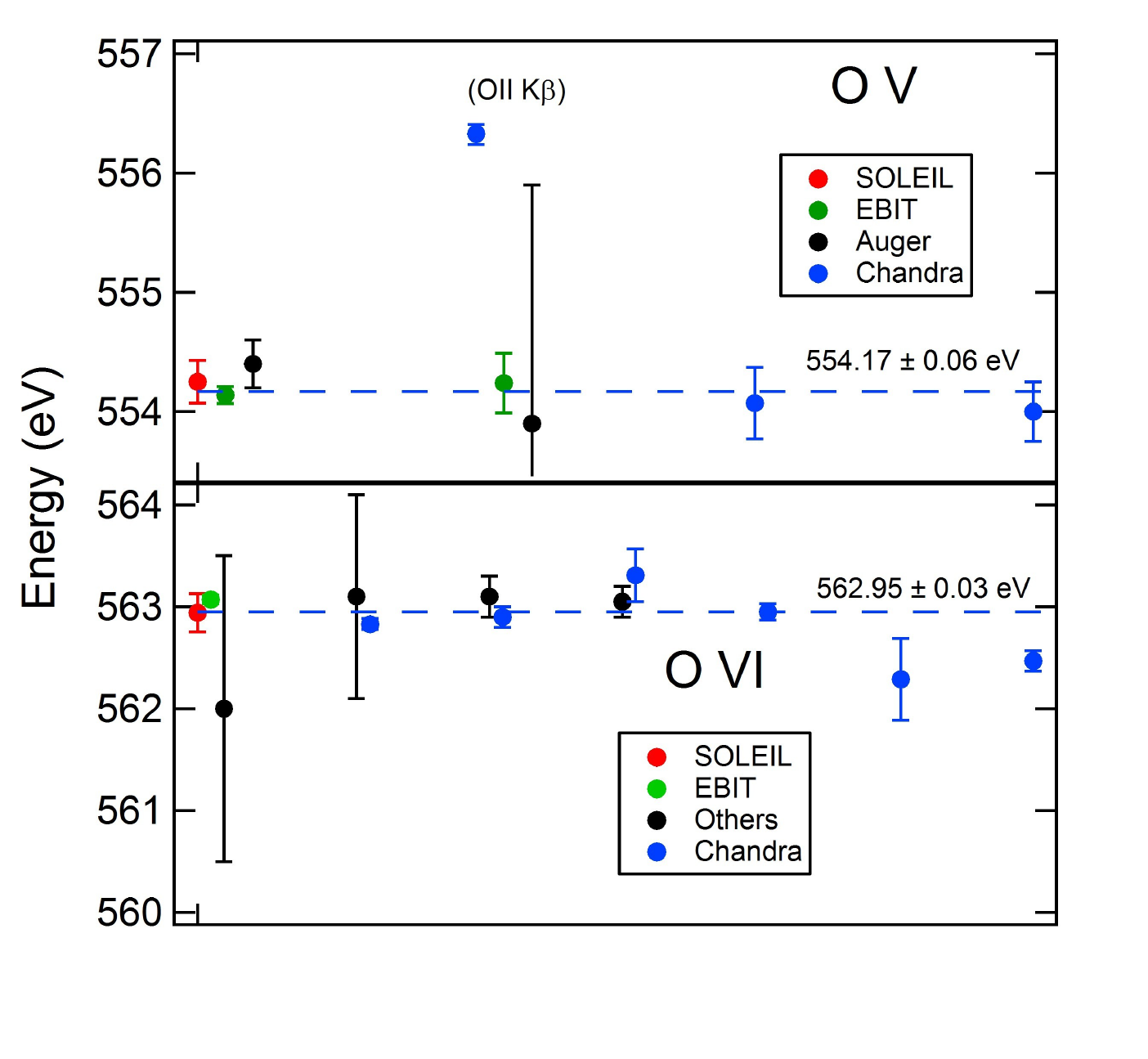}
\caption{\label{Figx6}(Colour online)  
            		Top panel, comparison of the $K_{\alpha}$ energies 
            		for O$^{4+}$ ions with values obtained from various 
		         experimental approaches, EBIT, Auger, 
		         the current SOLEIL measurements and Chandra observations.
	             Bottom panel,  similarly for 
	             the $K_{\alpha}$ energies for O$^{5+}$ ions 
	             with various experimental and observations
		         with the current SOLEIL measurements.
		         The dashed line (blue)
		         in both panels is the ponderated mean of all the values.}
\end{center}
\end{figure*}

Fig \ref{Figx6} illustrates more precisely the discrepancies 
for the O$^{4+}$ and O$^{5+}$ ions, $K_{\alpha}$ resonance lines 
between ground based experiments and prior satellite observations. 
The present results are for the O$^{4+}$(OV) ion (top panel) 
 and the O$^{5+}$(OVI) ion (bottom panel). 
 Table \ref{compare} quantifies the differences 
 between the ground based experimental measurements 
 (Synchrotron radiation (SR), and EBIT), and the 
 satellite observations (Chandra, and XMM-Newton).  
 We note in passing that the measurements 
 performed at two independent ground based light source
 experiments highlighted discrepancies in 
 the energies of the $K_{\alpha}$ lines along the 
 atomic oxygen iso-nuclear sequence compared to satellite observations. 
 Furthermore, we point out that the Chandra observations 
 of Liao and co-workers \citep{Liao2013},
 particularly  for the OV line at 556.33 eV 
 appears to be a misidentification. This line 
 is instead identified as the OII $K_{\beta}$ line. 
 We have indicated this in Fig.\ref{Figx6}.  This mis-identification 
 was indicated in the observations of 
Gatuzz and co-workers \citep{Gatuzz2014} and confirmed 
by the definitive observations of Nicastro and 
 co-workers \citep{Nicastro2016a}, who found the 
 OII $K_{\beta}$ line at 556.482 $\pm$ 0.50 eV from the
 average of 29 x-ray sources.

The discrepancies with the satellite observations 
are as yet not fully understood \citep{Kallman2016}.
We note, for the synchrotron measurements, 
the energy calibration was originally 
based on Electron Energy Loss Spectroscopy (EELS) 
measurements of 
O$_2$ $\rm 1s \rightarrow \pi^{\ast}$ 
performed by \cite{Brion1980}  
and ion yield measurements of 
O$_2$ $\rm 1s \rightarrow 3 \sigma$ 
by \cite{Tanaka2008}.
More recently the energy calibration 
is based on O$_2$ K-shell absorption 
spectra and Photo Electron 
Spectroscopy (PES) for $\rm 2s$ removal 
in Neon \citep{Bizau2015}. As can clearly 
be seen from Table \ref{calibrate} 
the difference between the two types of
energy calibrations is the order of 0.22 eV, 
and therefore cannot account for the larger energy
differences with the satellite observations.
Further independent ground based synchrotron light sources measurements 
would be of great beneficial help to try and minimise this source 
of calibration error.

\section{Conclusions}\label{secConclusions}
For the first time high-resolution photoinisation cross-section 
measurements have been made on Be-like and Li-like 
atomic oxygen ions in the vicinity of their strong $\rm 1s \rightarrow 2p$ 
 resonances and in the $\rm 1s \rightarrow np$ resonant 
region for O$^{4+}$ ions.  The measurements are compared with 
theoretical results from the R-matrix approach, the SCUNC method, 
and with available satellite observations.
Resonance features present in the spectra and predicted by
the R-matrix with pseudo-states (RMPS) method show excellent 
agreement  with the measurements made at the 
SOLEIL radiation facility. A detailed comparison 
of  our results (for both systems) 
are in agreement with predictions from 
other similar highly sophisticated theoretical approaches. 
 Within the R-matrix approach, we note that both basis sets and models 
 gave suitable results compared to experiment.  
The collision models with basis set B 
gave the more favourable results compared to experiment.

A detailed analysis of the resonance parameters 
indicates some differences  with the present 
SOLEIL experimental results, particularly for the weaker resonance 
strengths in the O$^{4+}$ spectrum. We speculate that this 
may be due to the lack of electron correlation included in our 
theoretical model.  
We have highlighted various discrepancies with previous 
satellite observations. 
Overall the theoretical resonance positions 
are in suitable agreement with 
current ground based SOLEIL experimental measurements.   
The theoretical cross-section data has been 
benchmarked against high resolution measurements  
and as such would be suitable to be 
incorporated into databases such as 
CLOUDY \citep{Ferland1998,Ferland2003}, XSTAR \citep{Kallman2001} 
and AtomDB \citep{Foster2012} that are widely 
used for astrophysical modelling.

\section*{Acknowledgments}
The authors would like to thank the SOLEIL staff,
in particular, J. Bozek and S. Nandi
for their helpful assistance during the measurements. 
We thank F. Nicastro for the provision of the
Chandra spectra for the blazar Mkn 421 and
 illuminating discussions on the differences with 
 the Chandra observations.
T. R. Kallman, at Nasa Goddard,
J. C. Raymond and R. K. Smith at the 
Harvard-Smithsonian Center for Astrophysics 
are also thanked for discussions 
on the astrophysical applications.
 B MMcL acknowledges support from the U.S. 
 National Science Foundation through a 
grant to ITAMP at the Harvard-Smithsonian Center 
for Astrophysics, the RTRA network Triangle de la Physique 
and  Queen's University Belfast for the award 
of a Visiting Research Fellowship (VRF). MFG 
acknowledges Qatar University for funding support 
through the startup grant No. QUSG-CAS-DMSP-14/15-4. 
The $R$-matrix computational work was  performed 
at the National Energy Research 
Scientific Computing Center (NERSC) in Berkeley, 
CA, USA  and at the High Performance 
Computing Center  Stuttgart (HLRS) of the 
University of Stuttgart, Stuttgart, Germany. 
Grants of computational time at the National 
Energy Research Scientific Computing Center 
in Berkeley, CA, USA and at the High Performance 
Computing Center Stuttgart (HLRS) 
of the University of Stuttgart, Stuttgart, 
Germany are gratefully acknowledged.
\bibliographystyle{mnras}                       
\bibliography{oions}

\begin{thebibliography}{}
\makeatletter
\relax
\def\mn@urlcharsother{\let\do\@makeother \do\$\do\&\do\#\do\^\do\_\do\%\do\~}
\def\mn@doi{\begingroup\mn@urlcharsother \@ifnextchar [ {\mn@doi@}
  {\mn@doi@[]}}
\def\mn@doi@[#1]#2{\def\@tempa{#1}\ifx\@tempa\@empty \href
  {http://dx.doi.org/#2} {doi:#2}\else \href {http://dx.doi.org/#2} {#1}\fi
  \endgroup}
\def\mn@eprint#1#2{\mn@eprint@#1:#2::\@nil}
\def\mn@eprint@arXiv#1{\href {http://arxiv.org/abs/#1} {{\tt arXiv:#1}}}
\def\mn@eprint@dblp#1{\href {http://dblp.uni-trier.de/rec/bibtex/#1.xml}
  {dblp:#1}}
\def\mn@eprint@#1:#2:#3:#4\@nil{\def\@tempa {#1}\def\@tempb {#2}\def\@tempc
  {#3}\ifx \@tempc \@empty \let \@tempc \@tempb \let \@tempb \@tempa \fi \ifx
  \@tempb \@empty \def\@tempb {arXiv}\fi \@ifundefined
  {mn@eprint@\@tempb}{\@tempb:\@tempc}{\expandafter \expandafter \csname
  mn@eprint@\@tempb\endcsname \expandafter{\@tempc}}}

\bibitem[\protect\citeauthoryear{{Al Shorman} et~al.,}{{Al Shorman}
  et~al.}{2013}]{Soleil2013}
{Al Shorman} M.~M.,  et~al., 2013, {J. Phys. B: At. Mol. Opt. Phys.},
  \textbf{46}, 195701

\bibitem[\protect\citeauthoryear{{Badnell}}{{Badnell}}{1986}]{Badnell1986}
{Badnell} N.~R.,  1986, {J. Phys. B: At. Mol. Opt. Phys.}, \textbf{19}, 382

\bibitem[\protect\citeauthoryear{{Badnell}}{{Badnell}}{2011}]{Badnell2011}
{Badnell} N.~R.,  2011, {Comput. Phys. Commun.}, \textbf{182}, 1528

\bibitem[\protect\citeauthoryear{{Ballance} \& {Griffin}}{{Ballance} \&
  {Griffin}}{2006}]{Ballance2006}
{Ballance} C.~P.,  {Griffin} D.~C.,  2006, J. Phys. B: At. Mol. Opt. Phys.,
  \textbf{39}, 3617

\bibitem[\protect\citeauthoryear{{Behar}, {Rasmuseeen}, {Blustin}, {Sako},
  {Kahn}, {Kaastra}, {Branduardi-Raymont}  \& {Steenbrugge}}{{Behar}
  et~al.}{2003}]{Behar2003}
{Behar} E.,  {Rasmuseeen} A.~P.,  {Blustin} A.~J.,  {Sako} M.,  {Kahn} S.~M.,
  {Kaastra} J.~S.,  {Branduardi-Raymont} G.,   {Steenbrugge} K.~C.,  2003,
  {Astrophys. J.}, \textbf{598}, 232

\bibitem[\protect\citeauthoryear{{Berkowitz}}{{Berkowitz}}{1979}]{berko1979}
{Berkowitz} J.,  1979, Photoabsorption, Photoionization and Photoelectron
  Spectroscopy.
Academic Press, New York, USA

\bibitem[\protect\citeauthoryear{{Berrington}, {Quigley}  \&
  {Zhang}}{{Berrington} et~al.}{1997}]{Berrington1997}
{Berrington} K.,  {Quigley} L.,   {Zhang} H.~L.,  1997, {J. Phys. B: At. Mol.
  \& Phys.}, \textbf{30}, 5409

\bibitem[\protect\citeauthoryear{{Bingcong} \& {Wensheng}}{{Bingcong} \&
  {Wensheng}}{2000}]{BW2000}
{Bingcong} G.,  {Wensheng} D.,  2000, {Phys. Rev. A}, \textbf{62}, 032705

\bibitem[\protect\citeauthoryear{{Bizau}, {Cubaynes}, {Guilbaud}, {Al Shorman},
  {Gharaibeh}, {Ababneh}, {Blancard}  \& {McLaughlin}}{{Bizau}
  et~al.}{2015}]{Bizau2015}
{Bizau} J.-M.,  {Cubaynes} D.,  {Guilbaud} S.,  {Al Shorman} M.~M.,
  {Gharaibeh} M.~F.,  {Ababneh} I.~Q.,  {Blancard} C.,   {McLaughlin} B.~M.,
  2015, {Phys. Rev. A.}, \textbf{92}, 023401

\bibitem[\protect\citeauthoryear{{Bizau} et~al.,}{{Bizau}
  et~al.}{2016}]{Bizau2016}
{Bizau} J.~M.,  et~al., 2016, J. Elec. Spec. Relat. Phenom., \textbf{210}, 5

\bibitem[\protect\citeauthoryear{{Blustin}, {Branduardi-Raymont}, {Behar},
  {Kaastra}, {Kahn}, {Page}, {Sako}  \& {Steenbrugge}}{{Blustin}
  et~al.}{2002}]{Blustin2002}
{Blustin} A.~J.,  {Branduardi-Raymont} G.,  {Behar} E.,  {Kaastra} J.~S.,
  {Kahn} S.~M.,  {Page} M.~J.,  {Sako} M.,   {Steenbrugge} K.~C.,  2002,
  {Astron. and Astrophys.}, \textbf{392}, 453

\bibitem[\protect\citeauthoryear{{Blustin} et~al.,}{{Blustin}
  et~al.}{2003}]{Blustin2003}
{Blustin} A.~J.,  et~al., 2003, {Astron. and Astrophys.}, \textbf{403}, 481

\bibitem[\protect\citeauthoryear{{Bruch}, {Schmeider}, {Schwarz}, {Meinhart},
  {Johnson}  \& {Taulberg}}{{Bruch} et~al.}{1979}]{Bruch1979}
{Bruch} R.,  {Schmeider} D.,  {Schwarz} W. H.~E.,  {Meinhart} M.,  {Johnson}
  B.~M.,   {Taulberg} K.,  1979, {Phys. Rev. A}, \textbf{19}, {\it 587}

\bibitem[\protect\citeauthoryear{{Bruch}, {Stolerfohr}, {Datz}, {Miller},
  {Pepmiller}, {Yamazaki}, {Krause}  \& {Swenson}}{{Bruch}
  et~al.}{1987}]{Bruch1987}
{Bruch} R.,  {Stolerfohr} N.,  {Datz} S.,  {Miller} P.~D.,  {Pepmiller} P.~L.,
  {Yamazaki} Y.,  {Krause} H.~F.,   {Swenson} J.~K.,  1987, {Phys. Rev. A},
  \textbf{35}, {\it 4114}

\bibitem[\protect\citeauthoryear{{Burke}}{{Burke}}{2011}]{Burke2011}
{Burke} P.~G.,  2011, R-Matrix Theory of Atomic Collisions: Application to
  Atomic, Molecular and Optical Processes.
Springer, New York, USA

\bibitem[\protect\citeauthoryear{{Cabot}, {Wang}  \& {Yao}}{{Cabot}
  et~al.}{2013}]{cabot2013}
{Cabot} S. H.~C.,  {Wang} Q.~D.,   {Yao} Y.,  2013, {Mon. Not. Roy. Ast. Soc.},
  \textbf{431}, 511

\bibitem[\protect\citeauthoryear{{Cassinelli}, {Waldron}, {Sanders}, {Harnden
  Jr}, {Rosner}  \& {Vaiana}}{{Cassinelli} et~al.}{1981}]{Cassinelli1981}
{Cassinelli} J.~P.,  {Waldron} W.~L.,  {Sanders} W.~T.,  {Harnden Jr} F.~R.,
  {Rosner} R.,   {Vaiana} G.~S.,  1981, {Astrophys. J}, \textbf{250}, 677

\bibitem[\protect\citeauthoryear{{Chen}}{{Chen}}{1985}]{Chen1985}
{Chen} M.~H.,  1985, {Phys. Rev. A}, \textbf{35}, 4579

\bibitem[\protect\citeauthoryear{{Chen}}{{Chen}}{1986}]{Chen1986}
{Chen} M.~H.,  1986, {At. Data Nucl. Data Tables}, \textbf{34}, 301

\bibitem[\protect\citeauthoryear{{Chen} \& {Crasemann}}{{Chen} \&
  {Crasemann}}{1987a}]{Chen1987}
{Chen} M.~H.,  {Crasemann} B.,  1987a, {Phys. Rev. A}, \textbf{35}, 4579

\bibitem[\protect\citeauthoryear{{Chen} \& {Crasemann}}{{Chen} \&
  {Crasemann}}{1987b}]{CC1987}
{Chen} M.~H.,  {Crasemann} B.,  1987b, {At. Data Nucl. Data Tables},
  \textbf{37}, 419

\bibitem[\protect\citeauthoryear{{Chen}, {Zhang}  \& {Gou}}{{Chen}
  et~al.}{2006}]{Chen2006}
{Chen} F.,  {Zhang} M.,   {Gou} B.,  2006, {J. Phys. B: At. Mol. Opt. Phys.},
  \textbf{39}, 4249

\bibitem[\protect\citeauthoryear{{Davis} \& {Chung}}{{Davis} \&
  {Chung}}{1985}]{Davis1985}
{Davis} B.~F.,  {Chung} K.,  1985, {Phys. Rev. A}, \textbf{31}, 3017

\bibitem[\protect\citeauthoryear{{Davis} \& {Chung}}{{Davis} \&
  {Chung}}{1989}]{Davis1989}
{Davis} B.~F.,  {Chung} K.~T.,  1989, Phys. Rev. A, 39, 3942

\bibitem[\protect\citeauthoryear{{Fano} \& {Cooper}}{{Fano} \&
  {Cooper}}{1968}]{Fano1968}
{Fano} U.,  {Cooper} J.~W.,  1968, Rev. Mod. Phys., \textbf{40}, 441

\bibitem[\protect\citeauthoryear{{Ferland}}{{Ferland}}{2003}]{Ferland2003}
{Ferland} G.~J.,  2003, Ann. Rev. of Astron. Astrophys., \textbf{41}, 517

\bibitem[\protect\citeauthoryear{{Ferland}, {Korista}, {Verner}, {Ferguson},
  {Kingdon}  \& {Verner}}{{Ferland} et~al.}{1998}]{Ferland1998}
{Ferland} G.~J.,  {Korista} K.~T.,  {Verner} D.~A.,  {Ferguson} J.~W.,
  {Kingdon} J.~B.,   {Verner} E.~M.,  1998, Pub. Astron. Soc. Pac. ({\it
  PASP}), \textbf{110}, 761

\bibitem[\protect\citeauthoryear{{Foster}, {Ji}, {Smith}  \&
  {Brickhouse}}{{Foster} et~al.}{2012}]{Foster2012}
{Foster} A.~R.,  {Ji} L.,  {Smith} R.~K.,   {Brickhouse} N.~S.,  2012,
  Astrophys. J, \textbf{756}, 128

\bibitem[\protect\citeauthoryear{{Gabriel}}{{Gabriel}}{1972}]{Gabriel1972}
{Gabriel} A.~H.,  1972, {Mon. Not. Roy. Astron. Soc.}, \textbf{160}, 99

\bibitem[\protect\citeauthoryear{{Garcia}, {Mendoza}, {Bautista}, {Gorczyca},
  {Kallman}  \& {Palmeri}}{{Garcia} et~al.}{2005}]{Garcia2005}
{Garcia} J.,  {Mendoza} C.,  {Bautista} M.~A.,  {Gorczyca} T.~W.,  {Kallman}
  T.~R.,   {Palmeri} P.,  2005, {Astrophys. J}, \textbf{779}, 78

\bibitem[\protect\citeauthoryear{{Gatuzz} et~al.,}{{Gatuzz}
  et~al.}{2013a}]{Gatuzz2013a}
{Gatuzz} E.,  et~al., 2013a, {Astrophys. J}, \textbf{768}, 60

\bibitem[\protect\citeauthoryear{{Gatuzz} et~al.,}{{Gatuzz}
  et~al.}{2013b}]{Gatuzz2013b}
{Gatuzz} E.,  et~al., 2013b, {Astrophys. J}, \textbf{778}, 83

\bibitem[\protect\citeauthoryear{{Gatuzz}, {Garcia}, {Mendoza}, {Kallman},
  {Bautius }  \& {Gorczyca}}{{Gatuzz} et~al.}{2014}]{Gatuzz2014}
{Gatuzz} E.,  {Garcia} J.,  {Mendoza} C.,  {Kallman} T.~R.,  {Bautius } M.~A.,
   {Gorczyca} T.~W.,  2014, {Astrophys. J}, \textbf{790}, 131

\bibitem[\protect\citeauthoryear{{Gorczyca} et~al.,}{{Gorczyca}
  et~al.}{2013}]{Gorczyca2013}
{Gorczyca} T.~W.,  et~al., 2013, {Astrophys. J}, \textbf{779}, 78

\bibitem[\protect\citeauthoryear{{Gu}, {Scmidt}, {Biersdorfer}, {Chen},
  {Thorn}, {{Tr\"{a}bert}}, {Behar}  \& {Kahn}}{{Gu} et~al.}{2005}]{Gu2005}
{Gu} M.~F.,  {Scmidt} M.,  {Biersdorfer} P.,  {Chen} H.,  {Thorn} D.~B.,
  {{Tr\"{a}bert}} E.,  {Behar} E.,   {Kahn} S.~M.,  2005, Astrophys. J.,
  \textbf{627}, 1066

\bibitem[\protect\citeauthoryear{{Gupta}, {Mathur}, {Krongold}, {Nicastro}  \&
  {Galeazzi}}{{Gupta} et~al.}{2012}]{Gupta2012}
{Gupta} A.,  {Mathur} S.,  {Krongold} Y.,  {Nicastro} F.,   {Galeazzi} M.,
  2012, Astrophys. J., 597, L8

\bibitem[\protect\citeauthoryear{{Gupta}, {Mathur}, {Galeazzi}  \&
  {Krongold}}{{Gupta} et~al.}{2014}]{Gupta2014}
{Gupta} A.,  {Mathur} S.,  {Galeazzi} M.,   {Krongold} Y.,  2014, Astrophys.
  Space Sci., 352, 775

\bibitem[\protect\citeauthoryear{{Hibbert}}{{Hibbert}}{1975}]{Hibbert1975}
{Hibbert} A.,  1975, {Comput. Phys. Commun.}, \textbf{9}, 141

\bibitem[\protect\citeauthoryear{{Hitchcock} \& {Brion}}{{Hitchcock} \&
  {Brion}}{1980}]{Brion1980}
{Hitchcock} A.~P.,  {Brion} C.~E.,  1980, J. Elec. Spec. Relat. Phenom.,
  \textbf{18}, 1

\bibitem[\protect\citeauthoryear{{Hoffmann}, {{M\"{u}ller}}, {Tinschert}  \&
  {Salzborn}}{{Hoffmann} et~al.}{1990}]{Hoffmann1990}
{Hoffmann} G.,  {{M\"{u}ller}} A.,  {Tinschert} K.,   {Salzborn} E.,  1990, Z.
  Phys. D - Atoms, Molecules and Clusters, \textbf{16}, 113

\bibitem[\protect\citeauthoryear{{Hoffmann}, {{M\"{u}ller}}, {Weiessbeeker},
  {Stenke}, {Tinschert}  \& {Salzborn}}{{Hoffmann} et~al.}{1991}]{Hoffmann1991}
{Hoffmann} G.,  {{M\"{u}ller}} A.,  {Weiessbeeker} K.,  {Stenke} M.,
  {Tinschert} K.,   {Salzborn} E.,  1991, Z. Phys. D - Atoms, Molecules and
  Clusters, \textbf{21}, S189

\bibitem[\protect\citeauthoryear{{Juett}, {Schulz}  \& {Chakrabarty}}{{Juett}
  et~al.}{2004}]{Juett2004}
{Juett} A.~M.,  {Schulz} N.~S.,   {Chakrabarty} D.,  2004, {Astrophys. J.},
  \textbf{612}, 308

\bibitem[\protect\citeauthoryear{{Kaastra}, {Steenbrugge}, {Raasen}, {van der
  Meer}, {Brinkman}, {Liedahl}, {Behar}  \& {de Rosa}}{{Kaastra}
  et~al.}{2002}]{Kaastra2002}
{Kaastra} J.~S.,  {Steenbrugge} K.,  {Raasen} A. J.~J.,  {van der Meer} R.
  L.~J.,  {Brinkman} A.~C.,  {Liedahl} D.~A.,  {Behar} E.,   {de Rosa} A.,
  2002, {Astron. and Astrophys.}, \textbf{386}, 427

\bibitem[\protect\citeauthoryear{{Kaastra} et~al.,}{{Kaastra}
  et~al.}{2004}]{Kaastra2004}
{Kaastra} J.~S.,  et~al., 2004, {Astron. and Astrophys.}, \textbf{428}, 57

\bibitem[\protect\citeauthoryear{{Kallman}}{{Kallman}}{2016}]{Kallman2016}
{Kallman} T.~R.,  2016, ~, private communication

\bibitem[\protect\citeauthoryear{{Kallman} \& {Bautista}}{{Kallman} \&
  {Bautista}}{2001}]{Kallman2001}
{Kallman} T.~R.,  {Bautista} M.~A.,  2001, Astrophys. J. Suppl. Ser.,
  \textbf{134}, 139

\bibitem[\protect\citeauthoryear{{Kjeldsen}, {Kristensen}, {Brooks}, {Folkman},
  {Knudsen}  \& {Andersen}}{{Kjeldsen} et~al.}{2002}]{Kjeldsen2002}
{Kjeldsen} H.,  {Kristensen} B.,  {Brooks} R.~L.,  {Folkman} H.,  {Knudsen} H.,
    {Andersen} T.,  2002, {Astrophys. J. Suppl. Ser.}, \textbf{138}, 219

\bibitem[\protect\citeauthoryear{{Krongold}, {Nicastro}, {Brickhouse}, {Elvis},
  {Liedahl}  \& {Mathur}}{{Krongold} et~al.}{2003}]{Krongold2003}
{Krongold} Y.,  {Nicastro} F.,  {Brickhouse} N.~S.,  {Elvis} M.,  {Liedahl}
  D.~A.,   {Mathur} S.,  2003, Astrophys. J., 597, 832

\bibitem[\protect\citeauthoryear{{Lee}, {Ogle}, {Canizares}, {Marshall},
  {Schulz}, {Morales}, {Fabian}  \& {Iwasawa}}{{Lee} et~al.}{2001}]{Lee2001}
{Lee} J.~C.,  {Ogle} P.~M.,  {Canizares} C.~R.,  {Marshall} H.,  {Schulz}
  N.~S.,  {Morales} R.,  {Fabian} A.~C.,   {Iwasawa} K.,  2001, {Astrophys.
  J.}, \textbf{554}, L13

\bibitem[\protect\citeauthoryear{{Liao}, {Zhang}  \& {Yao}}{{Liao}
  et~al.}{2013}]{Liao2013}
{Liao} J.~Y.,  {Zhang} S.-N.,   {Yao} Y.,  2013, {Astrophys. J}, \textbf{774},
  116

\bibitem[\protect\citeauthoryear{{Lin}, {Hsue}  \& {Chung}}{{Lin}
  et~al.}{2001}]{Lin2001}
{Lin} H.,  {Hsue} C.-S.,   {Chung} K.~T.,  2001, {Phys. Rev A}, \textbf{65},
  032706

\bibitem[\protect\citeauthoryear{{McLaughlin} \& {Ballance}}{{McLaughlin} \&
  {Ballance}}{2015}]{McLaughlin2015a}
{McLaughlin} B.~M.,  {Ballance} C.~P.,  2015, in {Resch} M.~M.,  {Kovalenko}
  Y.,  {Fotch} E.,  {Bez} W.,   {Kobaysah} H.,  eds, , Sustained Simulated
  Performance 2014.
Springer, Berlin, Germany, pp 173--190

\bibitem[\protect\citeauthoryear{{McLaughlin}, {Ballance}, {Bowen}, {Gardenghi}
   \& {Stolte}}{{McLaughlin} et~al.}{2013a}]{Stolte2013}
{McLaughlin} B.~M.,  {Ballance} C.~P.,  {Bowen} K.~P.,  {Gardenghi} D.~J.,
  {Stolte} W.~C.,  2013a, {Astrophys. J}, \textbf{771}, L8

\bibitem[\protect\citeauthoryear{{McLaughlin}, {Ballance}, {Bowen}  \&
  {Stolte}}{{McLaughlin} et~al.}{2013b}]{Oxygen2013}
{McLaughlin} B.~M.,  {Ballance} C.~P.,  {Bowen} K. P.~{Gardenghi} D.~J.,
  {Stolte} W.~C.,  2013b, {Astrophys. J}, \textbf{779}, L31

\bibitem[\protect\citeauthoryear{{McLaughlin}, {Bizau}, {Cubaynes}, {Al
  Shorman}, {Guilbaud}, {Sakho}, {Blancard}  \& {Gharaibeh}}{{McLaughlin}
  et~al.}{2014}]{Soleil2014}
{McLaughlin} B.~M.,  {Bizau} J.-M.,  {Cubaynes} D.,  {Al Shorman} M.~M.,
  {Guilbaud} S.,  {Sakho} I.,  {Blancard} C.,   {Gharaibeh} M.~F.,  2014, {J.
  Phys. B: At. Mol. Opt. Phys.}, \textbf{47}, 065201

\bibitem[\protect\citeauthoryear{{McLaughlin}, {Ballance}, {Pindzola}  \&
  {M{\"u}ller}}{{McLaughlin} et~al.}{2015}]{McLaughlin2015b}
{McLaughlin} B.~M.,  {Ballance} C.~P.,  {Pindzola} M.~S.,   {M{\"u}ller} A.,
  2015, in {Nagel} W.~E.,  {Kr\"{o}ner} D.~H.,   {Resch} M.~M.,  eds, , High
  Performance Computing in Science and Engineering'14.
Springer, Berlin,Germany, pp 23--40

\bibitem[\protect\citeauthoryear{{McLaughlin}, {Ballance}, {Pindzola},
  {Schippers}  \& {M{\"u}ller}}{{McLaughlin} et~al.}{2016}]{McLaughlin2016a}
{McLaughlin} B.~M.,  {Ballance} C.~P.,  {Pindzola} M.~S.,  {Schippers} S.,
  {M{\"u}ller} A.,  2016, in {Nagel} W.~E.,  {Kr\"{o}ner} D.~H.,   {Resch}
  M.~M.,  eds, , High Performance Computing in Science and Engineering'15.
Springer, Berlin,Germany, pp 51--74

\bibitem[\protect\citeauthoryear{{McLaughlin}, {Ballance}, {Pindzola},
  {Stancil}, {Schippers}  \& {M{\"u}ller}}{{McLaughlin}
  et~al.}{2017}]{McLaughlin2017}
{McLaughlin} B.~M.,  {Ballance} C.~P.,  {Pindzola} M.~S.,  {Stancil} P.~C.,
  {Schippers} S.,   {M{\"u}ller} A.,  2017, in {Nagel} W.~E.,  {Kr\"{o}ner}
  D.~H.,   {Resch} M.~M.,  eds, , High Performance Computing in Science and
  Engineering'16.
Springer, Berlin,Germany, p. submitted for publication

\bibitem[\protect\citeauthoryear{{Mendoza} et~al.,}{{Mendoza}
  et~al.}{2012}]{Mendoza2012}
{Mendoza} C.,  et~al., 2012, {The Reliability of atomic data used for oxygen
  abundance determinations, Conference on Mapping Oxygen in the Universe,
  Instituto de Astrofisica de Canarias, May 14-18}, \url
  {http://www.iac.es/congreso/oxygenmap/media/presentations/}

\bibitem[\protect\citeauthoryear{{Miller} \& {Bergman}}{{Miller} \&
  {Bergman}}{2015}]{Miller2015}
{Miller} M.~J.,  {Bergman} J.~N.,  2015, Astrophys. J., 800, 14

\bibitem[\protect\citeauthoryear{{Mitnik}, {Pindzola}, {Griffin}  \&
  {Badnell}}{{Mitnik} et~al.}{1999}]{mit99}
{Mitnik} D.~M.,  {Pindzola} M.~S.,  {Griffin} D.~C.,   {Badnell} N.~R.,  1999,
  J. Phys B: At. Mol. Opt. \& Phys., \textbf{32}, L479

\bibitem[\protect\citeauthoryear{{Moore}}{{Moore}}{1993}]{Moore1993}
{Moore} C.~E.,  1993, in {Gallagher} J.~W.,  ed., , {CRC Series in Evaluated
  Data in Atomic Physics}.
CRC Press, Boca Raton, FL, p.~399

\bibitem[\protect\citeauthoryear{{M{\"u}ller}}{{M{\"u}ller}}{2015}]{Mueller2015b}
{M{\"u}ller} A.,  2015, Phys. Scr., \textbf{5}, 054004

\bibitem[\protect\citeauthoryear{{M{\"u}ller} et~al.,}{{M{\"u}ller}
  et~al.}{2009}]{Mueller2009}
{M{\"u}ller} A.,  et~al., 2009, J. Phys. B: At. Mol. Opt. Phys., \textbf{42},
  235602

\bibitem[\protect\citeauthoryear{{M{\"u}ller} et~al.,}{{M{\"u}ller}
  et~al.}{2010}]{Mueller2010}
{M{\"u}ller} A.,  et~al., 2010, J. Phys. B: At. Mol. Opt. Phys., \textbf{43},
  135602

\bibitem[\protect\citeauthoryear{{M{\"u}ller} et~al.,}{{M{\"u}ller}
  et~al.}{2014}]{Mueller2014b}
{M{\"u}ller} A.,  et~al., 2014, J. Phys. B: At. Mol. Opt. Phys., \textbf{47},
  135201

\bibitem[\protect\citeauthoryear{{Murakami}, {Safronova}  \& {Kato}}{{Murakami}
  et~al.}{2002}]{Safronova2002}
{Murakami} I.,  {Safronova} U.~I.,   {Kato} T.,  2002, Can. J. Phys.,
  \textbf{80}, 1525

\bibitem[\protect\citeauthoryear{{Nahar}, {Pradhan}  \& {Zhang}}{{Nahar}
  et~al.}{2001}]{Nahar2001}
{Nahar} S.~N.,  {Pradhan} A.~K.,   {Zhang} H.~L.,  2001, Phys. Rev. A, 63,
  060701(R)

\bibitem[\protect\citeauthoryear{{Nicastro}, {Senatore}, {Gupta}, {Guainazzi},
  {Mathur}, {Krongold}, {Elvis}  \& {Piro}}{{Nicastro}
  et~al.}{2016a}]{Nicastro2016a}
{Nicastro} F.,  {Senatore} F.,  {Gupta} A.,  {Guainazzi} M.,  {Mathur} S.,
  {Krongold} Y.,  {Elvis} M.,   {Piro} L.,  2016a, Mon. Not. Roy. Astro. Soc.,
  \textbf{457}, 676

\bibitem[\protect\citeauthoryear{{Nicastro}, {Senatore}, {Gupta}, {Guainazzi},
  {Mathur}, {Krongold}, {Elvis}  \& {Piro}}{{Nicastro}
  et~al.}{2016b}]{Nicastro2016b}
{Nicastro} F.,  {Senatore} F.,  {Gupta} A.,  {Guainazzi} M.,  {Mathur} S.,
  {Krongold} Y.,  {Elvis} M.,   {Piro} L.,  2016b, Mon. Not. Roy. Astro. Soc.,
  \textbf{458}, L123

\bibitem[\protect\citeauthoryear{{Nicastro}, {Senatore}, {Krongold}  \&
  {Elvis}}{{Nicastro} et~al.}{2016c}]{Nicastro2016c}
{Nicastro} F.,  {Senatore} F.,  {Krongold} Y.,   {Elvis} M.,  2016c, Astrophys.
  J., \textbf{828}, L12

\bibitem[\protect\citeauthoryear{{Nicolosi} \& {Tondello}}{{Nicolosi} \&
  {Tondello}}{1997}]{Tondello1977}
{Nicolosi} P.,  {Tondello} G.,  1997, J. Opt. Soc. Am., \textbf{67}, 1033

\bibitem[\protect\citeauthoryear{{Ogle}, {Mason}, {Page}, {Salvi}, {Cordova},
  {McHardy}  \& {Priedhorsky}}{{Ogle} et~al.}{2004}]{Ogle2004}
{Ogle} P.~M.,  {Mason} K.~O.,  {Page} M.~J.,  {Salvi} N.~J.,  {Cordova} F.~A.,
  {McHardy} I.~M.,   {Priedhorsky} W.~C.,  2004, {Astrophys. J}, \textbf{606},
  151

\bibitem[\protect\citeauthoryear{{Petrini} \& {Tully}}{{Petrini} \&
  {Tully}}{1991}]{Petrini1991}
{Petrini} D.,  {Tully} J.~A.,  1991, {Astron. Astrophys.}, \textbf{241}, 327

\bibitem[\protect\citeauthoryear{{Piangos} \& {Nicolaides}}{{Piangos} \&
  {Nicolaides}}{1993}]{Nicolaides1993}
{Piangos} N.~A.,  {Nicolaides} A.,  1993, {Phys. Rev. A}, \textbf{48}, 4142

\bibitem[\protect\citeauthoryear{{Pinto}, {Kaastra}, {Costantini}  \& {de
  Vries}}{{Pinto} et~al.}{2013}]{Pinto2013}
{Pinto} C.,  {Kaastra} J.~S.,  {Costantini} E.,   {de Vries} C.,  2013,
  {Astron. and Astrophys.}, \textbf{551}, 25

\bibitem[\protect\citeauthoryear{{Pinto}, {Costantini}, {Fabian}, {Kaastra}  \&
  {{in'tZand}}}{{Pinto} et~al.}{2014}]{Pinto2014}
{Pinto} C.,  {Costantini} E.,  {Fabian} A.~C.,  {Kaastra} J.~S.,   {{in'tZand}}
  J. J.~M.,  2014, {Astron. and Astrophys.}, \textbf{563}, A115

\bibitem[\protect\citeauthoryear{{Pradhan}}{{Pradhan}}{2000}]{Pradhan2000}
{Pradhan} A.~K.,  2000, Astrophys. J., \textbf{545}, L165

\bibitem[\protect\citeauthoryear{{Pradhan}, {Chen}, {Delahaye}, {Nahar}  \&
  {Oelgoetz}}{{Pradhan} et~al.}{2003}]{Pradhan2003}
{Pradhan} A.~K.,  {Chen} G.~X.,  {Delahaye} F.,  {Nahar} S.~N.,   {Oelgoetz}
  J.,  2003, {Mon. Not. R. Astron. Soc.}, \textbf{341}, 1268

\bibitem[\protect\citeauthoryear{{Ram\'irez}}{{Ram\'irez}}{2013}]{Ramirez2013}
{Ram\'irez} J.,  2013, {Astron. and Astrophys.}, \textbf{551}, A95

\bibitem[\protect\citeauthoryear{{Robicheaux}, {Gorczyca}, {Griffin},
  {Pindzola}  \& {Badnell}}{{Robicheaux} et~al.}{1995}]{damp}
{Robicheaux} F.,  {Gorczyca} T.~W.,  {Griffin} D.~C.,  {Pindzola} M.~S.,
  {Badnell} N.~R.,  1995, Phys. Rev. A, \textbf{52}, 1319

\bibitem[\protect\citeauthoryear{{Sakho}}{{Sakho}}{2011}]{Sakho2011}
{Sakho} I.,  2011, {Rad. Phys. Chem.}, \textbf{80}, 1295

\bibitem[\protect\citeauthoryear{{Sakho}}{{Sakho}}{2012}]{Sakho2012}
{Sakho} I.,  2012, {Phys. Rev. A}, \textbf{86}, 052511

\bibitem[\protect\citeauthoryear{{Sakho}, {Diop}, {Faye}, {S\`{e}ne}, {Gueye},
  {Ndao}  \& {Biaye}}{{Sakho} et~al.}{2013}]{Sakho2013}
{Sakho} I.,  {Diop} B.,  {Faye} M.,  {S\`{e}ne} A.,  {Gueye} M.,  {Ndao} A.~S.,
    {Biaye} M.~{Wagu\'{e}} A.,  2013, {At. Data. Nucl. Data Tables},
  \textbf{99}, 447

\bibitem[\protect\citeauthoryear{{Samson}}{{Samson}}{1967}]{Samson1967}
{Samson} J. A.~R.,  1967, {Techniques of Vacuum Ultraviolet Spectroscopy}.
John Wiley \& Sons, New York, USA

\bibitem[\protect\citeauthoryear{{Schmidt}, {Biersdorfer}, {Chen}, {Thorn},
  {{Tr\"{a}bert}}  \& {Behar}}{{Schmidt} et~al.}{2004}]{Schmidt2004}
{Schmidt} M.,  {Biersdorfer} P.,  {Chen} H.,  {Thorn} D.~B.,  {{Tr\"{a}bert}}
  E.,   {Behar} E.,  2004, Astrophys. J., \textbf{627}, 1066

\bibitem[\protect\citeauthoryear{{Scully} et~al.,}{{Scully}
  et~al.}{2005}]{Scully2005}
{Scully} S. W.~J.,  et~al., 2005, {J. Phys. B: At. Mol. Opt. Phys.},
  \textbf{38}, 1967

\bibitem[\protect\citeauthoryear{{Shore}}{{Shore}}{1967}]{Shore1967}
{Shore} B.~W.,  1967, Rev. Mod. Phys., \textbf{39}, 439

\bibitem[\protect\citeauthoryear{{Tanaka} et~al.,}{{Tanaka}
  et~al.}{2008}]{Tanaka2008}
{Tanaka} T.,  et~al., 2008, {Phys. Rev. A}, \textbf{78}, 022516

\bibitem[\protect\citeauthoryear{{West}}{{West}}{2001}]{West2001}
{West} J.,  2001, J. Phys. B: At. Mol. Opt. Phys., \textbf{34}, R45

\bibitem[\protect\citeauthoryear{{Wu} \& {Xi}}{{Wu} \& {Xi}}{1991}]{Wu1991}
{Wu} L.-J.,  {Xi} J.-H.,  1991, {J. Phys. B: At. Mol. Phys.}, \textbf{24}, 3351

\bibitem[\protect\citeauthoryear{{Yao}, {Shultz}, {Gu}, {Nowak}  \&
  {Canizares}}{{Yao} et~al.}{2009}]{Yao2009}
{Yao} Y.,  {Shultz} N.~S.,  {Gu} M.~F.,  {Nowak} M.~A.,   {Canizares} C.~R.,
  2009, {Astrophys. J}, \textbf{696}, 1418

\bibitem[\protect\citeauthoryear{{Zhang} \& {Yeager}}{{Zhang} \&
  {Yeager}}{2012a}]{Yeager2012b}
{Zhang} S.~B.,  {Yeager} D.~L.,  2012a, J. Mol. Structure, \textbf{1023}, 96

\bibitem[\protect\citeauthoryear{{Zhang} \& {Yeager}}{{Zhang} \&
  {Yeager}}{2012b}]{Yeager2012a}
{Zhang} S.~B.,  {Yeager} D.~L.,  2012b, Phys. Rev. A, \textbf{85}, 032515

\makeatother
\end{thebibliography}
%
% if your bibtex file is called example.bib
%%%%%%%%%%%%%%%%% APPENDICES %%%%%%%%%%%%%%%%%%%%%
%\appendix

% Don't change these lines
\bsp	
% typesetting comment

\label{lastpage}

\end{document}